# Overview of Beyond-CMOS Devices and A Uniform Methodology for Their Benchmarking


Dmitri E. Nikonov and Ian A. Young

Components Research, Intel Corp., MS RA3-252, 2501 NW 229th Ave., Hillsboro, OR

97124, email: dmitri.e.nikonov@intel.com


Ver. 3.4 09/21/2012

## Abstract


Multiple logic devices are presently under study within the Nanoelectronic Research Initiative (NRI) to carry the development of integrated circuits beyond the CMOS roadmap. Structure and operational principles of these devices are described. Theories used for benchmarking these devices are overviewed, and a general methodology is described for consistent estimates of the circuit area, switching time and energy. The results of the comparison of the NRI logic devices using these benchmarks are presented.


## 1. Introduction

The development of CMOS integrated circuits has had unprecedented success from the scaling of its dimensions each new technology generation. Its further development is charted over the next several years by the International Technology Roadmap for Semiconductors [1] (ITRS). From the ITRS projections it transpires that the scaling will be influenced by fundamental physical limits of device switching [2]. Due to this observation, research thrusts in the academia and the industry (most prominently, the Nanoelectronic Research Initiative - NRI) gained significant momentum towards demonstrating and thoroughly investigating feasible alternatives to CMOS.

The NRI group has also performed benchmarking of the "beyond-CMOS" devices [3]. Such an investigation is of utmost importance, since it permits identification and focusing of resources on researching the most promising devices. However this benchmarking investigation provided only a summary report of the results (e.g. Figure 1), without providing a uniform common methodology and details behind the benchmarking calculations needed to reproduce them. Also different investigators made sometimes widely different assumptions regarding the operating conditions and characteristics of individual devices. One of the main goals of this study was to obtain the values for the area, switching time, and switching energy of a set of standard circuits – an inverter with a fanout of 4, a 2-input NAND gate, and a 32-bit adder. We adopt the same goal in this paper just to be able to make a one-to-one comparison. Such benchmarks are very useful as they allowed for the first time a comparison of the promise of beyond-CMOS computing with the mainstream, CMOS, computing. However one needs to be aware of limitations of such an approach. It well may happen that a different circuit architecture



needs to be worked out for optimal performance of beyond-CMOS devices. For example, spintronic devices are non-volatile (preserve their state when the power is switched off), and a different circuit is needed to make use of its property to create normally-off instantly-on logic chips. Present benchmarks are not designed to measure this utility of devices. Moreover, different devices may prove to be best suited for different roles within a circuit. We conjecture that future integrated circuits will still contain a majority of CMOS devices with a few other beyond-CMOS devices performing various specialized functions. Also by focusing on the switching energy, one comprehends only the active power dissipation, but not the standby ("leakage") power. Such benchmarks need to be a subject of future research.

In this paper we are setting out to establish a consistent standard methodology for benchmarking beyond-CMOS logic devices in order to obtain a reliable set of metrics and fair comparison of these devices. In treatment of the devices we chose **simplicity** rather than rigor. Wherever possible, we used analytical expressions, rather than simulations, for calculations of benchmarks. At the present time the structure of the devices and their operational characteristics are not firmed up. We believe it presumptuous at this early stage of the research to expect the benchmarks to be accurate within less than a factor of 2. We strived for **uniformity** of benchmarking: same assumptions, relations, and schemes are applied to all devices to which they may pertain. Therefore the relative benefits of the devices under study might be more accurate than their absolute values. We also insisted on complete **transparency** of our benchmarks. All the equations and parameters used are listed in this paper. We provide the Matlab code used to generate all plots in this paper [4]. This way the readers can reproduce all the



results of the paper. Also they can plug in their values and assumptions and explore "what if" scenarios.

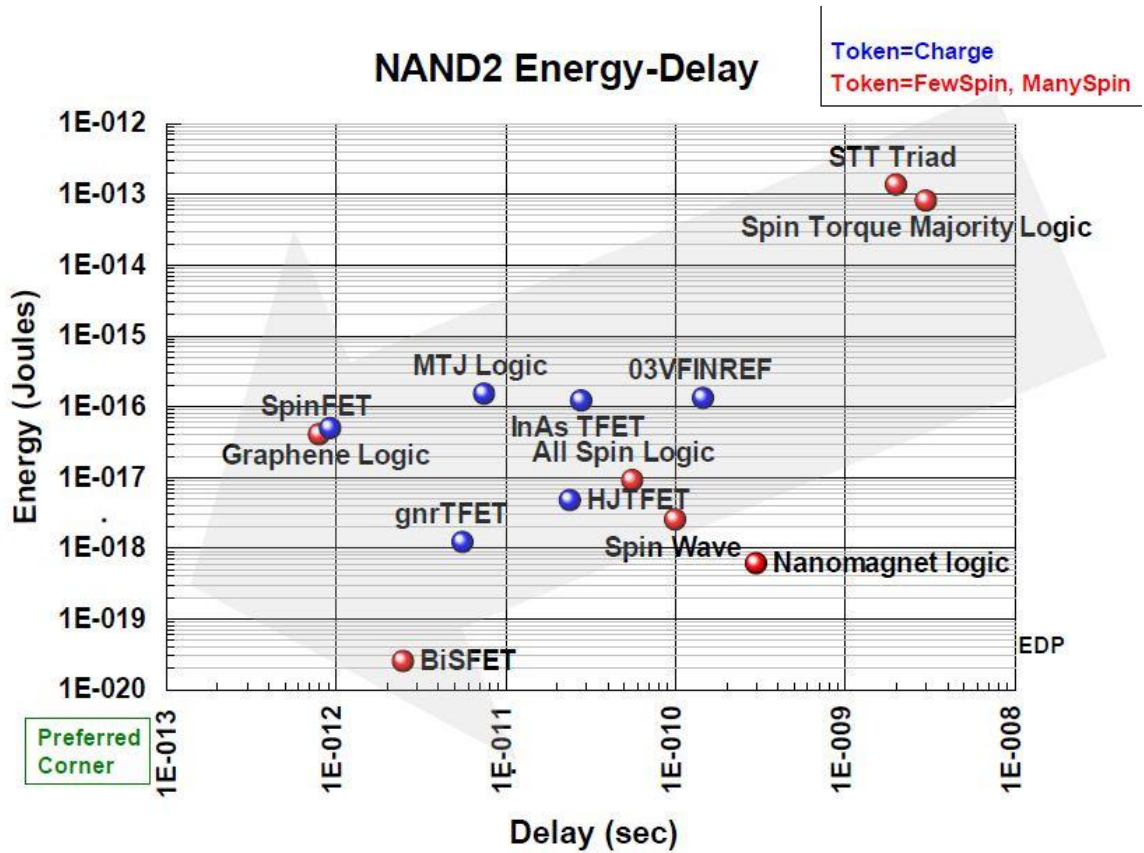

**Figure 1. Comparison of devices from K. Bernstein's presentation [5].**

The paper is structured as follows. The requirements of logic technologies are outlined in Section 2. Non-traditional computational variables are described in Section 3. The nomenclature of beyond-CMOS devices is introduced in section 4. Physical constants and device parameters are listed in Section 5. The principles of layout and assumptions about the gate size are described in Section 6. The method used for estimating the circuit area is explained in Section 7. General considerations for time and energy of capacitance charging are considered in Section 8. Simple analytical expressions for the switching time and energy of electronic circuits are proposed Section 9. Estimates



for various methods of magnetization switching are collected in Section 10. Section 11 contains the estimates of performance for spintronic devices. Overall benchmarks for devices are compared in Section 12. The consequences for the computational throughput and dissipated power are calculated in Section 13. The results of the paper are discussed in Section 14.

In this paper we limit the scope to only digital circuits. We do not consider analog, mixed or digital-to-analog conversion circuits. Moreover we only consider Boolean logic, thereby excluding non-Boolean [6] or neuromorphic [7] computing.

## 2. Tenets of logic and device interconnection

Before it can be considered as a candidate for an element of an integrated logic circuit, a solid-state device needs to satisfy a set of requirements [8], "logic tenets", such as:

    i.    Non-linear characteristics (related to noise margin and the signal-to-noise ratio)

    ii.    Power amplification (gain>1)

    iii.    Concatenation (output of one device can drive another)

    iv.    Feedback prevention (output does not affect input)

    v.    Complete set of Boolean operators (NOT, AND, OR, or equivalent)

In addition it needs to be competitive in the quantitative physical measures of:

    i.    Size (i.e. scalability)

    ii.    Switching time

    iii.    Switching energy (i.e. power dissipation)

And finally, technological requirements, such as

    i.    Room temperature or higher operation



ii.  Low sensitivity to parameters (e.g. fabrication variations)

iii.  Operational reliability

iv.  CMOS architectural compatibility (interface, connection scheme)

v.  CMOS process compatibility (fabricated on the same wafer)

vi.  Comprehending intrinsic and extrinsic parasitic and their interface to interconnect

may decide the fate of a device in the competition to be the next technology of choice.

Devices considered in this study are at various levels of technological maturity. Some

have been demonstrated experimentally, operation of others has been simulated, while

some are still at the concept stage. In this paper we do not review the status of

experimental work of the devices or discuss the issues of their manufacturability. Neither

are we trying to predict the success of their integration to logic circuits. We assume that

devices would work as they are intended and estimate their performance characteristics

starting from physical relations governing phenomena underlying them. We make

optimistic assumptions about the material parameters, lithography capabilities, and

device structures without pushing these values to their physical limits.

## 3. Computational variables and device classification

The beyond-CMOS devices encode information by various physical quantities, which we

call "computational variables". A list and a pictorial representation of types of

computational variables are shown in Figure 2. We use them to classify devices.

The first set of variables comprises charge, current, and voltage (designated as Q, I, and

V, respectively). It is very familiar to the readers, is the underlying group of variables in

electronic devices, which comprise the overwhelming majority of mainstream computing



and a few beyond-CMOS options. Ferroelectric devices are based on electric dipoles (designated as P). Ferroelectric transistors [8] have been researched and demonstrated for many years now, and we do not include them in this study. Spintronic [9] devices [10] rely on magnetic dipoles represented by ferromagnetic elements or electrons with polarized spins (designated as M). Orbitronic devices are the least understood ones, as they involve the orbital state (designated as Orb) of electrons in a molecule or a crystal, and sometimes a collective state of electrons, such as Bose condensate of excitons (designated as Cond). In the current study they are represented by a single device, the BisFET. There are other computational variables, described in the ITRS [1] ERD chapter, such as: mechanical position for NEMS devices, light intensity for photonic devices, timing of signals in neuromorphic computing, etc. But they are not used in this study.

**1. Charge, current, voltage (Q,I,V)**

**2. Electric dipole (P)**

**3. Magnetic dipole = spin (M)**

**4. Orbital state (Orb) in e.g. quantum well, molecule, crystal, also excitons, esp. Bose condensate (BC)**

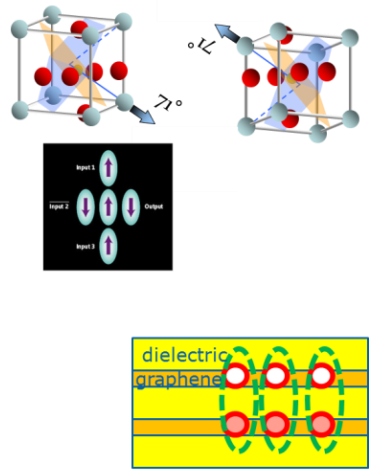

**Figure 2. Scheme of computational variables.**

The computational variables can have various roles in a logic circuit. To explain them, we envision a very abstract black-box diagram of a logic device shown in Figure 3. The computational variables encode the internal state, input and output signals and the controls for switching devices (including clocking).



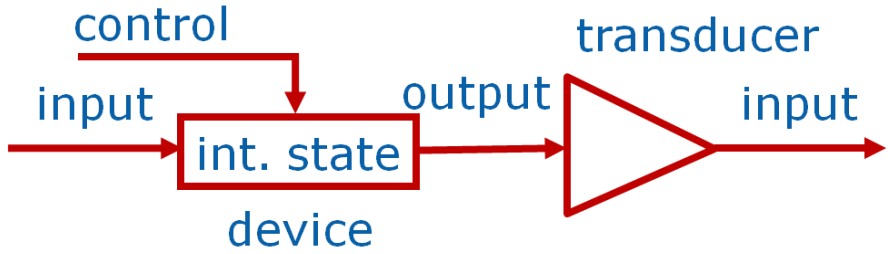



One of the tenets of logic (in Section 2) is that the output of one stage needs to be able to drive the input to the next stage. If this condition is not satisfied for a device, e.g. the input is a voltage signal but the output is a magnetization signal, special purpose devices which we call "transducers", are needed to convert one variable to another. The most important transducers discussed below are the ones converting electronic signals to spintronic ones and vice versa.

## 4. Nomenclature and short description of the devices

Devices are classified according to their computational variables as they represent inputs, outputs, and the internal state (Section 3) when the devices are connected in an integrated circuit, see Table 1. The subclass is assigned according to a phenomenon underlying the device operation. This is explained in more detailed as we go over devices one-by-one. We aim to give a very short description of a device features and nature of connections relevant for our discussion.

| Device name | acronym | input | control | int. state | output | class | subclass |
|---|---|---|---|---|---|---|---|
| Si MOSFET high performance | CMOS HP | V | Vg | Q | V | electronic | barrier |
| Si MOSFET low power | CMOS LP | V | Vg | Q | V | electronic | barrier |
| Homojunction TFET | HomJTFET | V | Vg | R | V | electronic | tunneling |
| Heterojunction TFET | HetJTFET | V | Vg | R | V | electronic | tunneling |
| Graphene nanoribbon TFET | gnrTFET | V | Vg | R | V | electronic | tunneling |
| Graphene pn-junction (Veselago) | GpnJ | V | Vg | R | V | electronic | refraction |
| Bilayer pseudospin FET | BisFET | V | Vg | BEC | V | orbitronic | exciton |
| SpinFET (Sughara-Tanaka) | SpinFET | V | Vg, Vm | Q, M | V | spintronic | spin drift |
| Spin torque domain wall | STT/DW | I | V | M | I | spintronic | domain wall |
| Spintronic majority gate | SMG | M | V | M | M | spintronic | domain wall |
| Spin torque triad | STTtriad | I | V | M | I | spintronic | nanomagnet |
| Spin torque oscillator | STOlogic | I | V | M | I | spintronic | nanomagnet |
| All spin logic device | ASLD | M | V | M | M | spintronic | spin diffusion |
| Spin wave device | SWD | M | I or V | M | M | spintronic | spin wave |
| Nanomagnetic logic | NML | M | B or V | M | M | spintronic | nanomagnet |

**Table 1. Nomenclature and classification of included devices.**

## 4.1 CMOS.

The complementary metal-oxide-semiconductor (CMOS) FET (Figure 4) is a familiar device. Its internal state is a charge on the capacitor of the gate dielectric. The input and gate voltages determine the output voltage (Figure 5). Switching is done by raising and lowering of the potential barrier for electrons in the channel due to a change of the gate voltage. For high performance **CMOS (CMOS HP)** we use the values from the 2011 edition of ITRS [1] PIDS chapter. There the technology node F=15nm, chosen for the NRI study, corresponds to the 2018 column. The low-power **CMOS (CMOS LP)** used in the NRI benchmarking study is envisioned as a low supply voltage (0.3V) device from [11]. It is to be noted that it is different from the low operating power (LOP CMOS) and low standby power (LSP CMOS) transistors considered in ITRS [1].



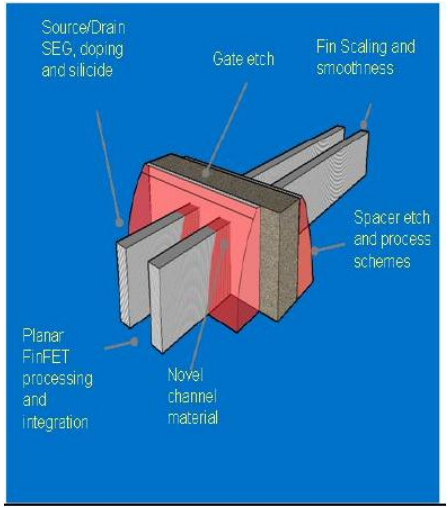

**Figure 4. Advanced Si multi-gate MOSFET according to [1]. The scheme is be used for both high-performance and low-power CMOS, depending on the applied and threshold voltages.**

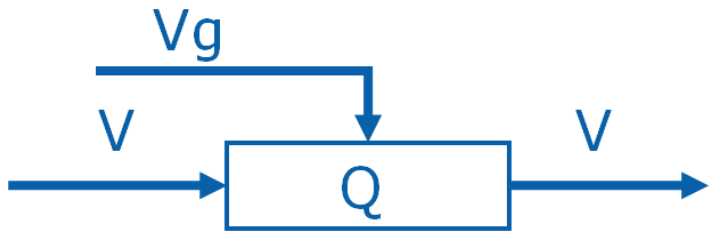

**Figure 5. Block-diagram for CMOS.**

### 4.2. Tunneling FET.

Tunneling field-effect transistors (TFET) [12] are considered under three material options – homojunction III-V material (HomJFET, Figure 6) specifically InAs double-gate transistor, heterojunction III-V material [13] (HetJFET, Figure 7) specifically InAs/GaSb double-gate transistor, and graphene nanoribbon (gnrFET, Figure 8). They have the same principle of operation and differ in performance parameters - supply voltage and drive current. Parameters for TFET are taken from simulation, such as [14]. Conduction in a TFET occurs through band-to-band tunneling (BTBT). Gate voltage shifts the bands in



energy and drastically changes the probabilities of tunneling. The block-diagram of TFET (Figure 10) is very similar to that of CMOS. Simulations show that charge on a gate of the TFET is smaller than the corresponding CMOS [15]. Therefore we prefer to associate the state of the device with the resistance of the channel (R). Besides, this smaller gate capacitance contributes to faster switching of circuits, an advantage of TFET beyond just comparing drive currents.

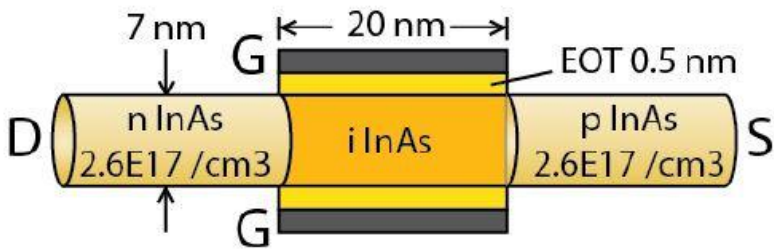

**Figure 6. III-V Homojunction Tunneling FET (HomJTFET)**

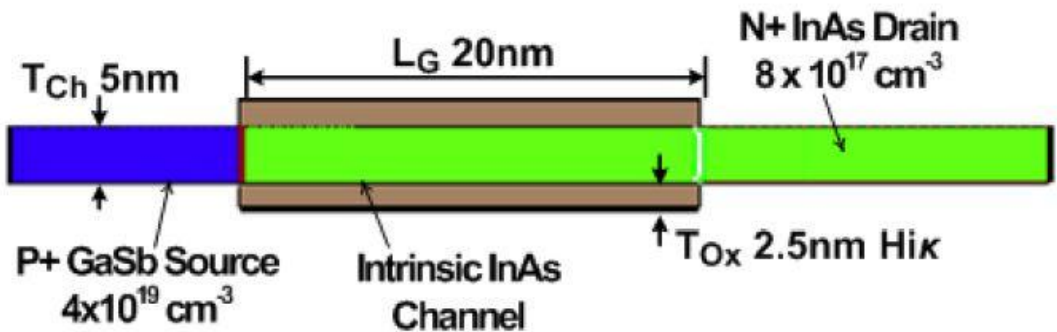

**Figure 7. Heterojunction (HetJTFET)**

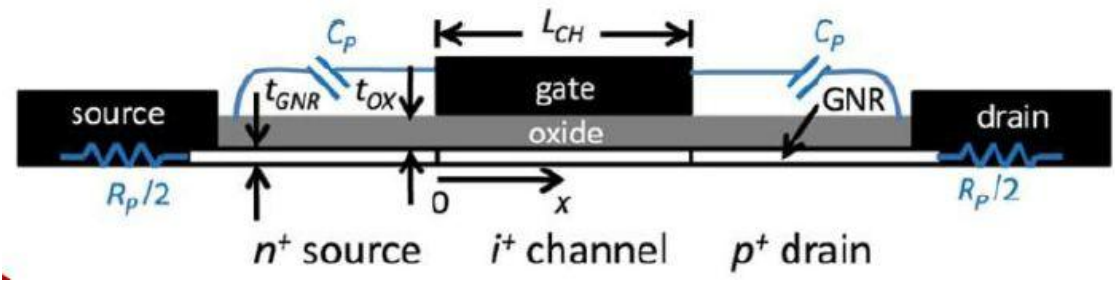

**Figure 8. Graphene nanoribbon (gnrTFET)**



### 4.3 Graphene pn-junction.

A graphene pn-junction (GpnJ, Figure 9) [16] device uses the junctions to switch the path of electrons. It shares the block-diagram with the TFET (Figure 10) signifying the change of the resistance as an internal state, but relies on a completely different physical phenomenon. Reflection of electrons from pn-junctions in graphene is highly dependent on the angle due to its peculiar bandstructure. Therefore, by switching the electrostatic p and n doping of graphene by applying voltage to electrodes, it is possible to achieve either a very high transmission or a total internal reflection of electrons. Thus current is directed to one vs. another output of the device.

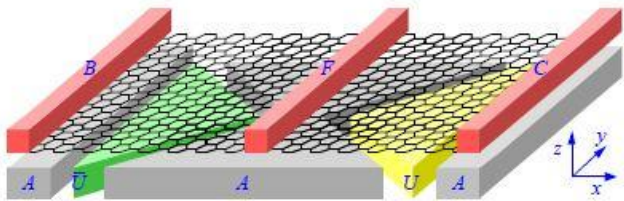

**Figure 9. Graphene pn-junction (GpnJ)**

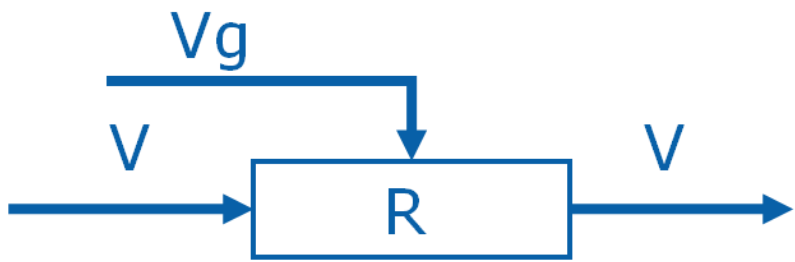

**Figure 10. Block-diagram for tunneling transistors and graphene pn-junctions.**

### 4.4 BisFET.

A bilayer pseudospin FET (BisFET, Figure 11) [17] is another graphene device. It exploits tunneling between two monolayers of graphene. Due to a stronger interaction of electrons and holes in graphene, it is expected that they will bind into excitons at a higher



temperature than in traditional semiconductors. If holes are injected into one monolayer and electrons into another monolayer, they may bind into excitons and these excitons might relax into a Bose-Einstein condensate (BEC) state. Considerable controversy exists about the critical temperature of BEC in bylayer graphene: the original proposal [18] suggest that it is above room temperature, while subsequent calculations [19] predict that it is much lower than 1K. The presence of BEC is expected to drastically increase the probability of tunneling due to its collective nature. BEC state can be destroyed due to changing the balance between electrons and holes by applying a gate voltage. Thus voltage controls the internal state (R) of the device related to the presence of the condensate, see Figure 12. The current between source (S) and drain (D) in Figure 11 is expected first to grow with the increase of voltage Vds and then decrease as the carrier imbalance destroys BEC (thus exhibiting negative differential resistance). The device proposal [17] postulates the current-voltage (I-V) curve with a peak at voltage of 5mV. For the present benchmarks we are using the results of quantum transport simulations [20,21,22] which exhibit I-V curves with peak voltages of 150mV to 400mV. Though they were performed for a different value of the interlayer coupling constant and different wiring of the device. The gates V(+) and V(-) are designed to maintain certain high densities of electrons and holes in the bottom and top graphene layers, respectively. These potentials need to be kept constant. Gate G covers only a half of the channel and is used to control the BisFETs in logic circuits. Its indended use is "current crowding": the applied voltage causes current to flow on one side of the channel. If current density there exceeds the peak value, conduction there drops, and the current is forced to the other side and it exceeds the peak current density there as well.



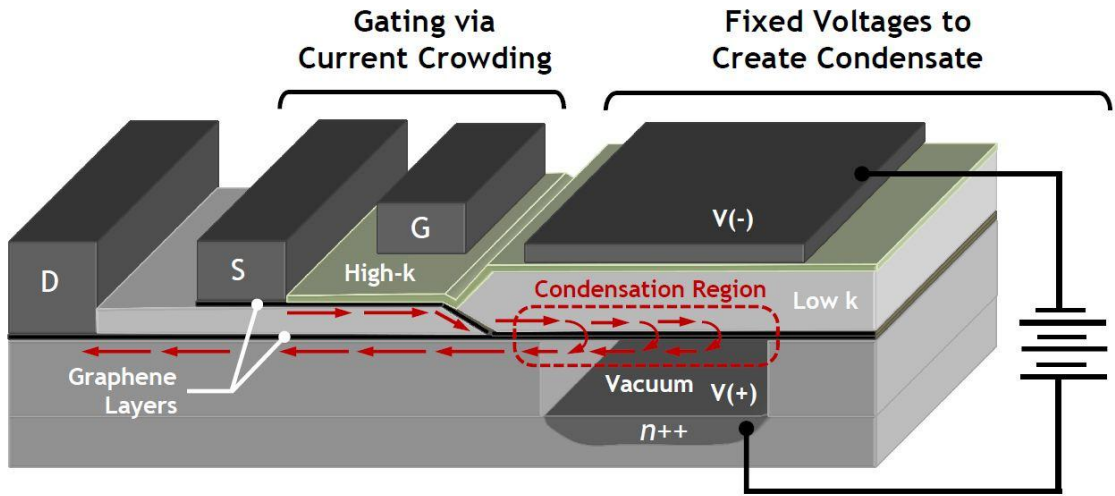

**Figure 11. Bilayer pseudospin (BisFET) scheme [23].**

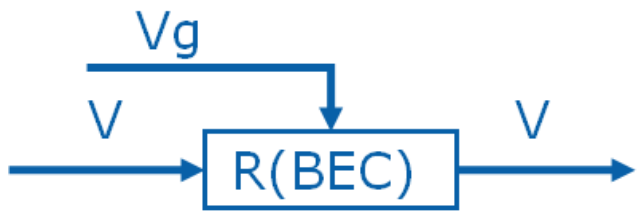

**Figure 12. Block-diagram for BisFET.**

## 4.5. SpinFET.

A spinFET (Figure 13) [24] combines a MOSFET and a switchable magnetic element. Its source and drain are made of ferromagnetic metals and another ferromagnet is positioned over a drain in order to detect its direction of magnetization via a tunneling magnetoresistance (TMR) effect. In addition to the usual FET functionality, R(Q), the resistance of a spinFET depends on the magnetization state, R(M), see Figure 14. If the magnetizations of the source and drain are parallel, the resistance of the channel is low. If they are anti-parallel, the resistance is high. In addition to the current through the



channel, the magnetization of the drain can be switched by a current from a terminal controlled by voltage Vm.

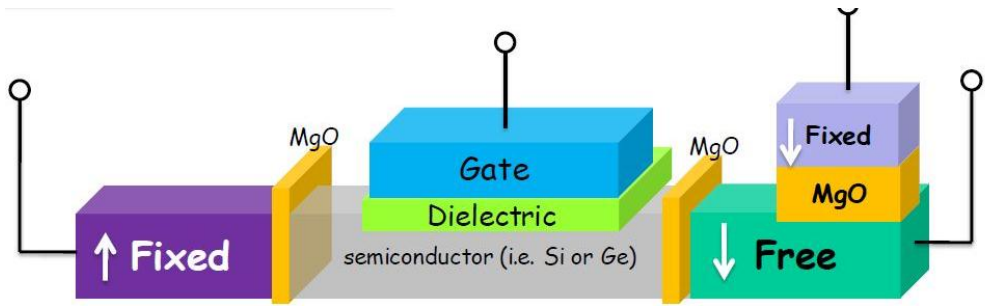

**Figure 13. SpinFET (Sughara-Tanaka) [25].**

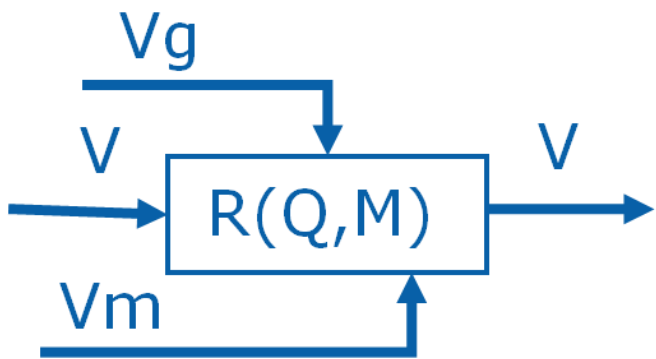

**Figure 14. Block-diagram for SpinFET.**

### 4.6. Spintronics features.

As in most of the spintronic devices, the Magnetic Tunnel Junction (MTJ) is required only at the output magnets in order to detect the direction of magnetization. MTJ is a stack of a ferromagnet, a tunneling dielectric, and another ferromagnet. It has a huge advantage in the values of magnetoresistance (MR) compared to spin valves, but at the price of around three orders of magnitude higher resistance-area product. A spin valve is a stack of a ferromagnet, a non-magnetic metal, and another ferromagnet. When only a



switching based on the spin torque principal is required, a spin valve is preferable and gives a comparable polarization of the spin current.

Most of the spintronic devices are non-volatile, i.e. their computational state is preserved even when the power to the circuit is turned off. The condition for non-volatility is stability of nanomagnets against thermal fluctuations. In other words, the magnet should have two states of equilibrium separated by a barrier with energy greater than $60k_BT$.

### 4.7. Spin transfer torque domain wall device.

A spin transfer torque / domain wall (STT/DW) device (Figure 15) [26] operates by motion of a domain wall in a ferromagnetic wire. The motion is caused by a spin transfer torque effect of a current along the wire (unlike a current perpendicular to the wire used in the rest of the spin torque devices in this paper). As the domain wall moves, the magnetization below the MTJ stack (in the middle of the device) switches, resulting in high or low resistance of the stack. Then a current from the "clock" terminal to the output through the MTJ is either high or low. The current is used to drive the inputs of next stages (Figure 22).



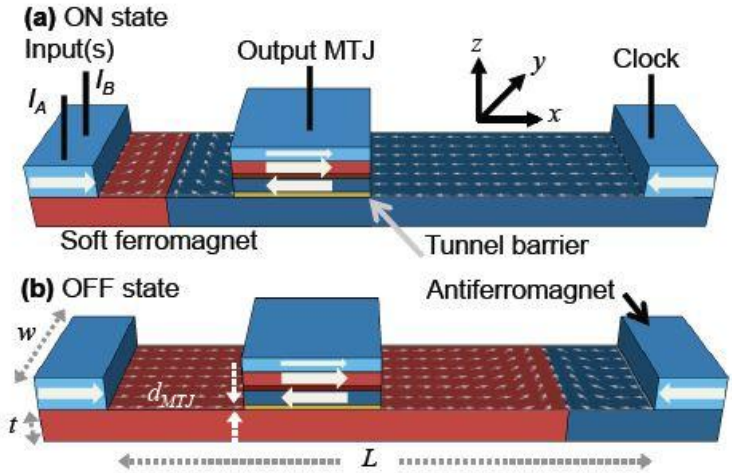

**Figure 15. Spin transfer torque domain wall logic (STT/DW). A domain wall separates regions with opposite magnetization (shown as red and blue).**

### 4.8. Spintronic majority gate.

A spintronic majority gate (SMG, Figure 16) [27,28] is implemented with a "cross" of ferromagnetic wires. It has three inputs and one output terminals formed over the ends of the "cross" as nanopillars with their own ferromagnetic layer. Current from each nanopillar exerts spin torque which aims to switch magnetization to a certain direction, depending on the sign of the current. The majority of the inputs win and enforce their direction of the magnetization. This is sensed via the Tunneling Magnetoresistance (TMR) effect using a sense amplifier (Figure 23). Like a few other spintronic devices, inputs can be switched and outputs can be sensed by a magnetoelectric (ME) cell (rather than spin torque and TMR), see Figure 24. For the principle of operation of a ME cell, see Section 10. It should be noted that electric-spin conversion (= writing and sensing of magnetization) does not need to occur at every majority gate input/output. Instead multiple majority gates can be cascaded into a larger magnetic circuit. For example, three



majority gates are enough to form a one-bit full adder. Electric-spin conversion then happens only at the adder interface between the magnetic and the electronic circuits.

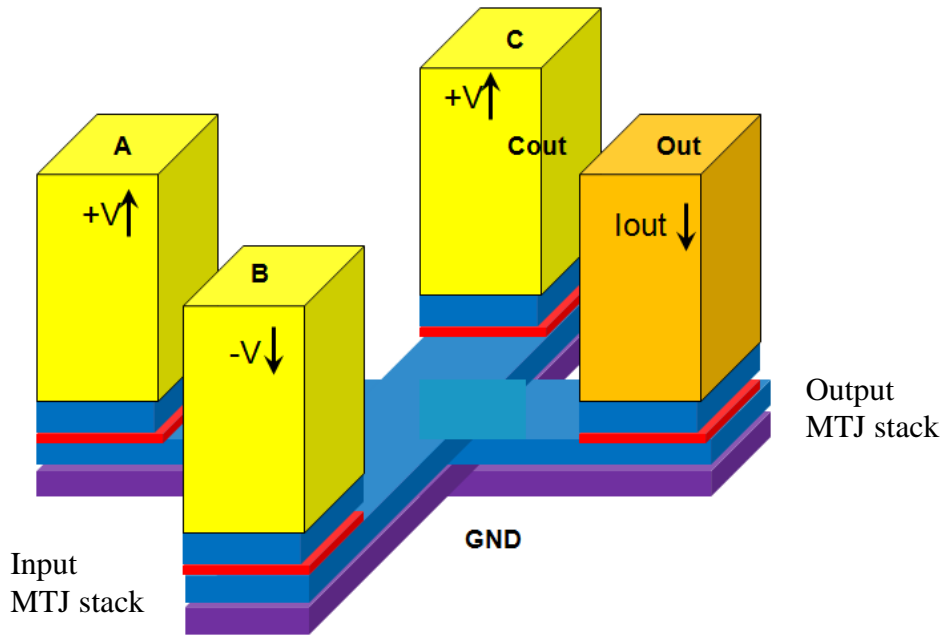

**Figure 16. Spintronic majority gate (SMG).**

**4.9 Spin transfer torque triad.**

A spin transfer torque triad (STTtriad) [29] element works in a manner similar to the SMG. However it consists of triangular structures (Figure 17) with two inputs and one output. Then the current from one triangle is used to drive spin torque switching in other triangles, in a manner similar to STT/DW (Figure 22). This geometry implies electric-spin conversion at every computing element, rather than cascading magnetic signals.



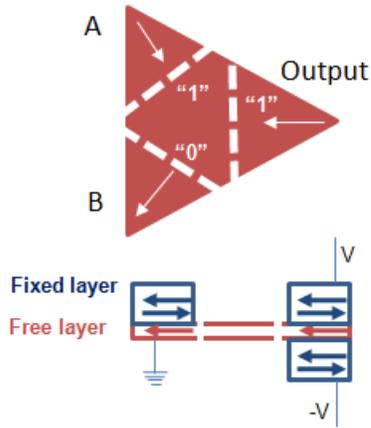

**Figure 17. Spin transfer torque triad (STTtriad) [29].**

**4.10. Spin torque oscillator logic.**

Spin torque oscillator (STO) logic [30] contains oscillators which are driven by spin torque from currents entering each of them through nanopillars on top (Figure 18). STOs are combined into a majority gate with three input and one output oscillator. The oscillators have a common ferromagnetic layer, similar to SMG. Oscillations cause spin waves to propagate in the common layer and thus the oscillators' signal couple. The frequency of the output oscillator is determined by the majority of inputs and serves as the logic signal in the circuit. By the nature of the signals, STO logic is described by the block diagram in Figure 22. Since in order to oscillate the STOs must have only one position of equilibrium, therefore this type of logic is volatile.



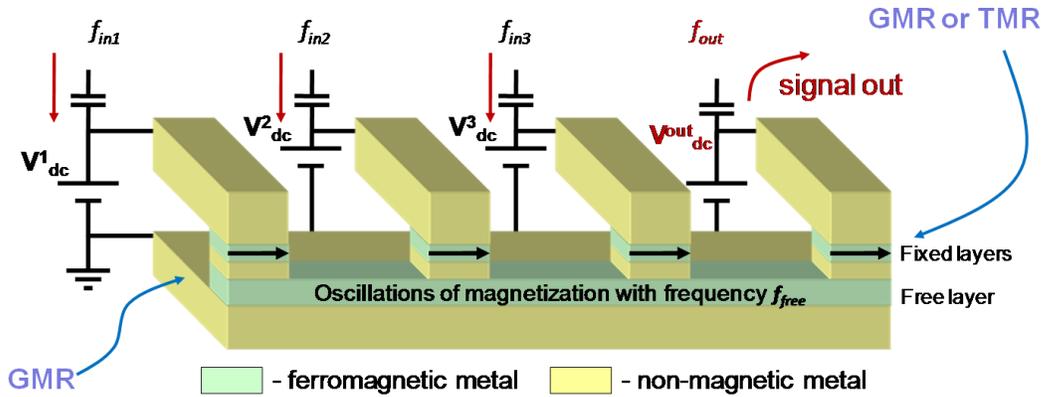

**Figure 18. Spin torque oscillator (STO) logic**

### 4.11. All spin logic.

An all spin logic device (ASLD) [31, 32] is formed by nanomagnets placed over a copper wire (Figure 19). Each nanomagnet has an input and the output sides separated by an insulator. Voltage supplied to the top of each nanomagnet drives a current to the ground terminal nearby. Due to this current spin polarized electrons accumulate near each nanomagnet. Concentrations of polarized spins are different at the input and output sides of the two neighboring magnets, which causes a diffusion spin current to occur. This spin current exerts torque on a nanomagnet and is able to switch its polarization. The nanomagnets are often arranged into majority gates. The nanomagnets can be concatenated according to the block-diagram in Figure 23, for example, to produce a one-bit full adder, as in Figure 19.



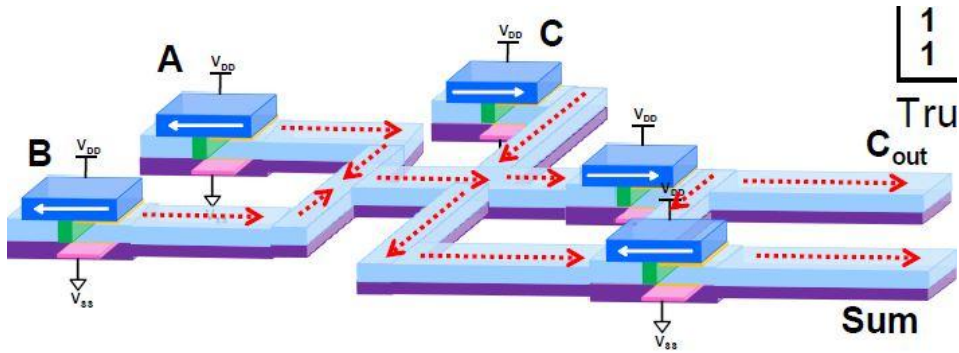

**Figure 19. All spin logic devices (ASLD) forming a one bit of a full adder [33].**

## 4.12. Spin wave device.

A spin wave device (SWD, Figure 20) [34, 35] contains nanomagnets connected by ferromagnetic wires. Spin waves are excited in the wires and propagate along them. Short pulses of spin waves containing a wide range of frequencies are used. There are two versions of the device in which: 1) spin waves are excited and detected by an RF antenna; this version does not need nanomagnets and is volatile, corresponding to the scheme in Figure 23; 2) spin waves are excited and detected by a magnetoelectric cell; it includes nanomagnets and is non-volatile, corresponding to the scheme in Figure 24. The magnetization of a nanomagnet determines the phase of the spin wave at a given coordinate. In its turn, this phase determines to what direction magnetization will switch in the output nanomagnet. SWD gates are readily amenable to cascading.



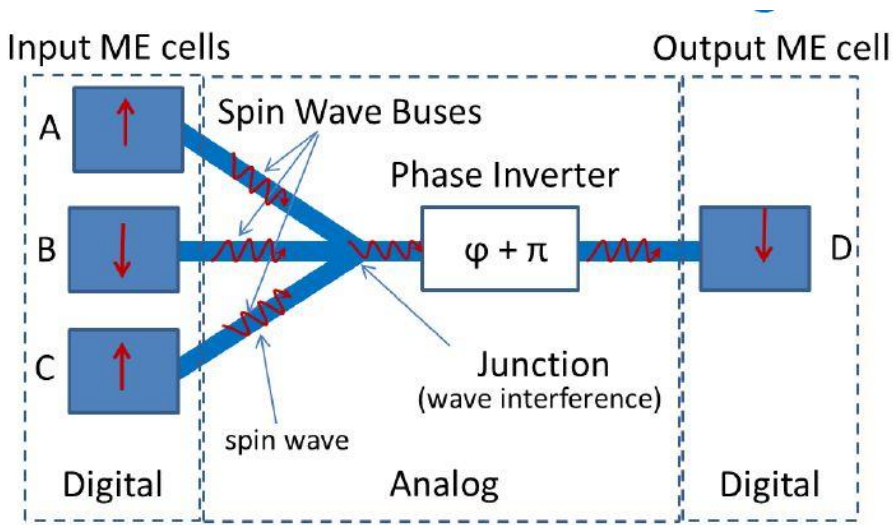

**Figure 20. Spin wave device (SWD).**

### 4.13. Nanomagnetic logic.

Nanomagnetic logic (NML, Figure 21) [36] consists of a chain of nanomagnets. They interact by magnetic dipole coupling. NML is the only device in the NRI suite of spintronic devices which uses Bennett clocking [36]. In other words, its operation occurs by preparing all nanomagnets in a quasi-stable equilibrium state by the action of a magnetic field from a current in a wire (current clocking, Figure 23) or effective magnetic field from the charging a magnetoelectric cell (voltage clocking, Figure 24). In both cases the state of magnets at the input causes the magnets in the circuit to choose one of the stable equilibrium states determined by the dipole interaction between magnets. It is easy to cascade NML gates.



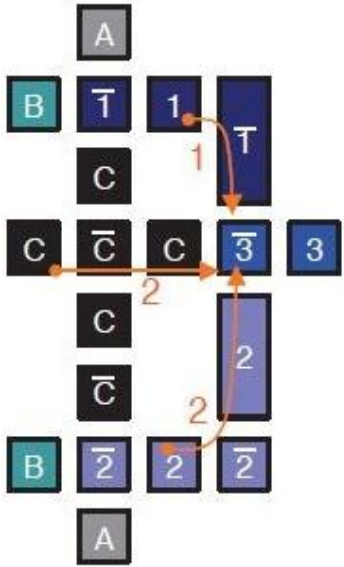

**Figure 21. Nanomagnetic logic (NML).**

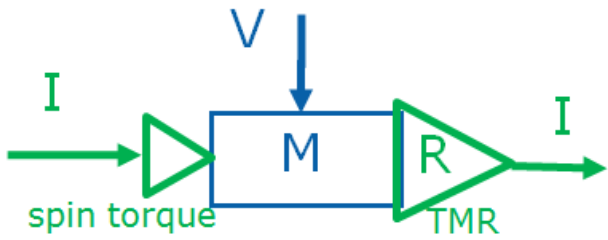

**Figure 22. Block-diagram for STT/DW, STOlogic, and STTtriad driven by spin torque.**

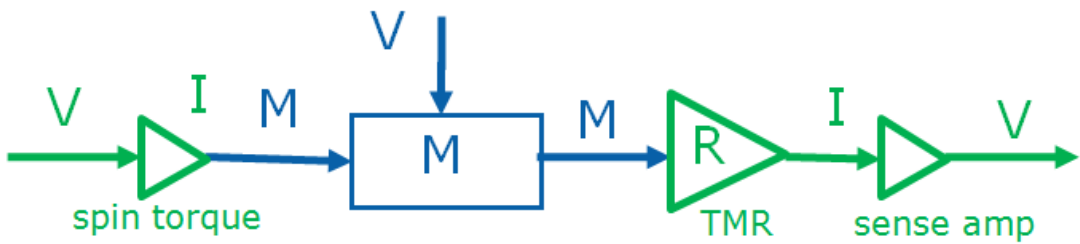

**Figure 23. Block-diagram for SMG, ASLD, SWD, and NML driven by spin torque.**



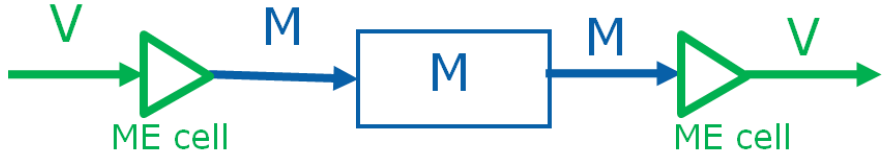

<span style="color:#4a7ebb">**Figure 24. Block-diagram for SMG, ASLD, SWD, and NML with magnetoelectric switching.**</span>

### 4.14. Other devices.

Beyond-CMOS devices previously considered within the NRI studies or in other research overviews that are not included in this study are: Datta-Das spin modulator [37], electronic ratchet [38]; graphene thermal logic [39]; SET/BDD [40]; electron structure modulation transistor [41]; RAMA [42,43]; resonant injection enhanced FET [44]; magnetic domain-wall Logic [45]; domain wall majority gate logic [46]; MottFET [47]; spin Hall effect transistor [48]; few spin device [49]; magnetic tunnel junction (MTJ) plus CMOS logic [50,51,52,53]. We also do not cover the optoelectronic excitonic transistor, which was demonstrated [54,55,56], as well as other proposed versions of excitonic FETs [3,57,58]. The reason for non-inclusion is either lack of input information to build benchmarks or insufficient research activity on this device at NRI.

## 5. Constants and parameters

In the estimates of the device performance we are using the following constants and parameters collected in Table 2 and Table 3.

.

| PHYSICAL QUANTITY | SYMBOL | UNITS | VALUE |
|---|---|---|---|



| Electron charge, absolute value | $e$ | C | 1.602176565e-19 |
|---|---|---|---|
| Electron mass | $m_e$ | kg | 9.10938291e-31 |
| Planck's constant | $\hbar$ | J*s | 1.054571726e-34 |
| Permittivity of vacuum | $\varepsilon_0$ | C$^2$/(J*m) | 8.854187817e-12 |
| Speed of light | $c$ | m/s | 2.99792458e8 |
| Boltzmann's constant | $k_B$ | J/K | 1.3806488e-23 |
| Permeability of vacuum | $\mu_0$ | kg*m/ C$^2$ | $(\varepsilon_0 c^2)^{-1}$ |
| Lande factor | $g$ | | 2 |
| Bohr magneton | $\mu_B$ | A*m$^2$ | $e\hbar/(2m_e)$ |
| Gyromagnetic constant | $\gamma$ | C/kg | $ge/(2m_e)$ |
| Quantum resistance | $R_q$ | $\Omega$ | $2\pi\hbar/e^2$ |

**Table 2. Fundamental physical constant used in the analysis.**

| PHYSICAL QUANTITY | SYMBOL | UNITS | TYPICAL VALUE |
|---|---|---|---|
| Ambient temperature | $T$ | K | 300 |
| Dielectric constant of SiO$_2$ | $\varepsilon$ | | 3.9 |
| Dielectric constant of interlayer dielectric | $\varepsilon_{ILD}$ | | 2 |
| Particle velocity in graphene | $v_F$ | m/s | 8e5 |
| Resistivity of copper | $\rho$ | $\Omega$*m | 1.9e-8 |
| Magnetization in a ferromagnet, in plane | $M_s$ | A/m | 1e6 |
| Magnetization in a ferromagnet, perpendicular | $M_{sp}$ | A/m | 0.3e6 |
| Injected spin polarization | $P$ | | 0.8 |
| Gilbert damping | $\alpha$ | | 0.01 |
| Perpendicular magnetic anisotropy | $K_u$ | J/m$^3$ | 6e5 |
| Current density for domain wall motion | $J_{dw}$ | A/m$^2$ | 1.4e11 |
| Magnetic permeability around wires | $\mu$ | | 4 |
| Magnetic induction from a wire, required | $B_{wi}$ | T | 0.01 |
| Multiferroic electric polarization [59] | $P_{mf}$ | C/m$^2$ | 0.55 |
| Multiferroic switching field [59] | $\mathbb{E}_{mf}$ | V/m | 2e7 |
| Multiferroic exchange bias | $B_{mf}$ | T | 0.03 |
| Magnetostrictive switching field [60] | $\mathbb{E}_{ms}$ | V/m | 6e5 |
| Magnetostrictive effective field [61] | $B_{ms}$ | T | 0.03 |
| Dielectric constant of piezoelectric [60] | $\varepsilon_{ms}$ | | 20000 |
| Magnetoelectric coefficient [62] | $\alpha_{me}$ | s/m | 5.6e-9 |
| Dielectric constant of magnetoelectric | $\varepsilon_{me}$ | | 1000 |

**Table 3. Material constants used in the analysis.**

## 6. Layout principles

In this section we propose a simple but general method to estimate the areas of beyond-CMOS circuits. The semiconductor process generations are labeled by



characteristic lithography size called the DRAM's half-pitch, $F$. It is set to 15nm in this study. ITRS [1] projects values of many device parameters depending on $F$. We use the following device parameters in Table 4.

| PHYSICAL QUANTITY | SYMBOL | UNITS | TYPICAL VALUE |
|---|---|---|---|
| Gate length | $L_g$ | m | 1.28e-8 |
| Equivalent oxide thickness, nominal | $EOT_0$ | m | 0.68e-9 |
| Equivalent oxide thickness, electrical | $EOT$ | m | 1.08e-9 |
| Contact resistance | $R_c$ | $\Omega$*m | 1.6e-4 |
| Contact capacitance | $C_c$ | F/m | 1.2e-10 |
| Magnetic induction to switch NML | $B_{nd}$ | T | 0.01 |
| Ferromagnet thickness | $d_{fm}$ | m | 2e-9 |
| Piezoelectric thickness for magnetoelectric effect | $d_{me}$ | m | 25e-9 |

**Table 4. Device parameters used in the analysis.**

The layout of the devices is governed by the design rules which specify minimum width and spacing of various elements in the circuit [63]. Scalable design rules are formulated in units of maximum mask misalignment $\lambda$, [63], where typically $\lambda = F/2$. An example of design rules is shown in Figure 25.



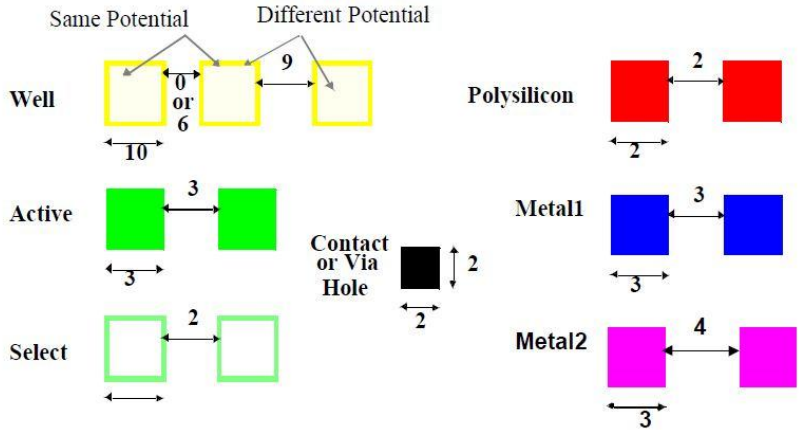

**Figure 25. Example of scalable design rules for a CMOS foundry.**

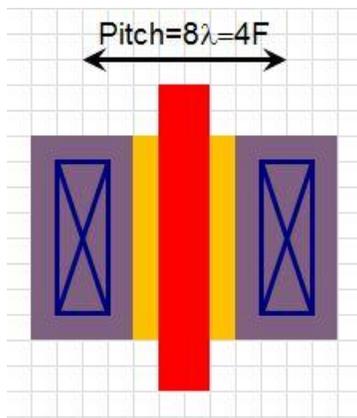

**Figure 26. Layout of a transistor, after [64].**

We see in Figure 26 that the pitch of metal-1 in the contacted transistor is $p_m = 8\lambda$.

On the other hand, the metal 1 contacted gate pitch scales with the process generation [65], see Figure 27.



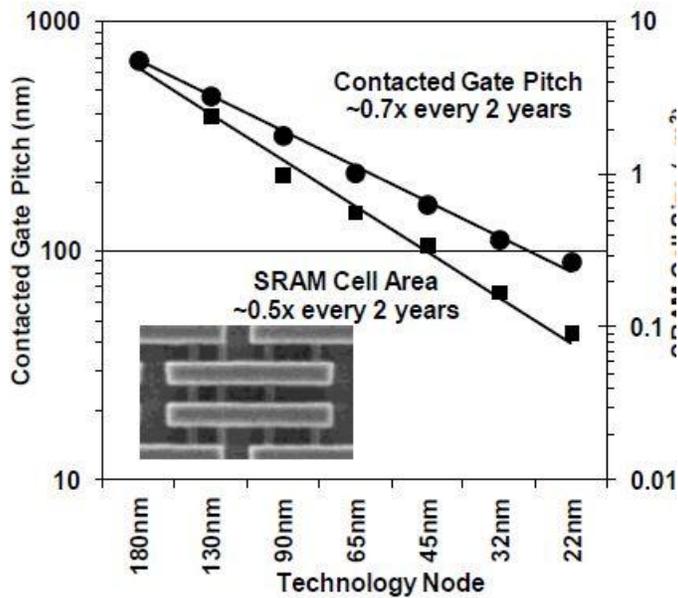



From this plot we approximately obtain that the metal 1 pitch is $p_m = 4F$. This confirms the previous relation $F = 2\lambda$. Since all the logic circuits (even spintronics circuits) need electric contacts at the terminals of gates, their size will be determined by the metallization pitch much more than by their intrinsic device sizes. We proceeds to draw layouts for simple circuits for all of the considered devices in the following section. We are using the color scheme shown in Figure 28 to designate mask layers. We notices that their area can be estimated by counting the pitches of the most important lines (diffusion, gate ("poly"), metal 1) [64] as shown in Figure 29. Here and in the rest of the document, one cell of the grid corresponds to size λ. The contacted pitch for all these important lines proves to be $8\lambda$. Thus in the rest of the paper we take the minimum pitch between any contacted circuit elements to be $8\lambda = 4F$.



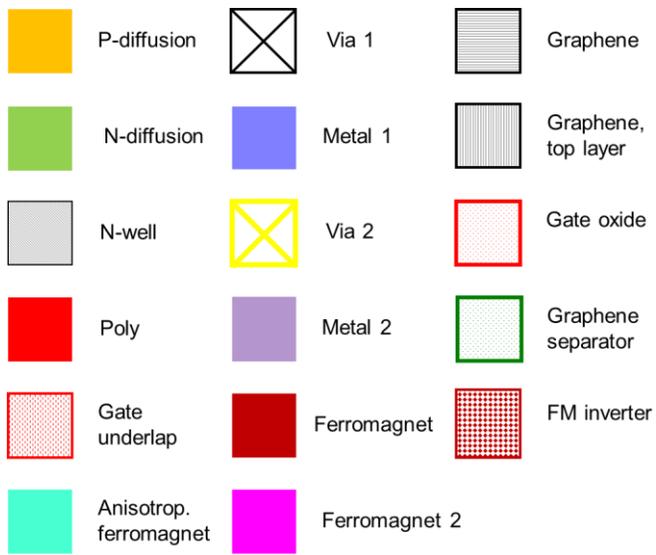

| | | | | |
|---|---|---|---|
| P-diffusion | Via 1 | Graphene |
| N-diffusion | Metal 1 | Graphene, top layer |
| N-well | Via 2 | Gate oxide |
| Poly | Metal 2 | Graphene separator |
| Gate underlap | Ferromagnet | FM inverter |
| Anisotrop. ferromagnet | Ferromagnet 2 | |

**Figure 28. Legend for the layout layers used in this paper.**

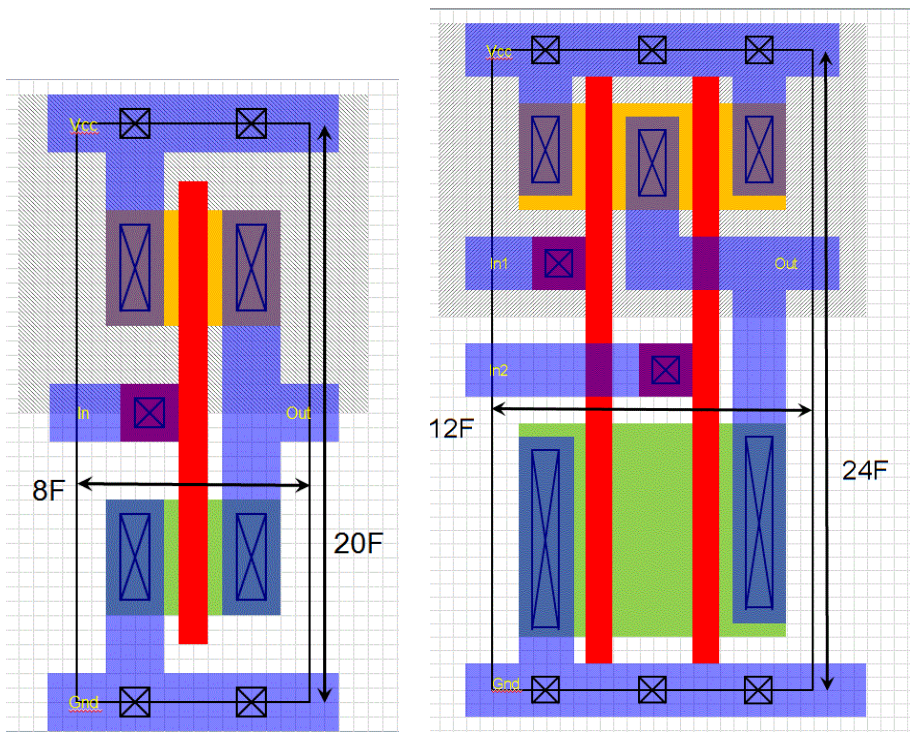

**Figure 29. Layout and area estimation for a CMOS inverter and a 2-input NAND gate.**

## 7. Circuit area estimation

We use the parameters in Table 5 to calculate the sizes of devices and circuits.



| PARAMETER | TYPICAL VALUE |
|---|---|
| Transistor width $w_X$ / $F$ | 4 |
| Spin contact width $w_S$ / $F$ | 1 |
| Width of metal-1 $w_m$ / $F$ | 2 |
| Height of metal-1 $p_m$ / $F$ | 4 |
| Separation of metal-1 lines $s_m$ / $F$ | 2 |
| Interlayer -1 thickness $d_d$ / $F$ | 4 |
| Inverter FO1 length and width / pitch | 2 x 5 |
| NAND2 length and width / pitch | 3 x 6 |
| XOR area / NAND2 area | 2 |
| Gate area overhead, $M_{gate}$ | 1.5 |
| Bit area overhead, $M_{bit}$ | 1.5 |
| Adder area overhead, $M_{add}$ | 1.5 |
| Interconnect length / pitch | 5 |

**Table 5. Geometry parameters of the devices.**

The size of the intrinsic device (a transistor-like device has subscript $X$, and a spintronic

device has subscript $S$) is obtained as

$$a_{\text{int}} = \left( w_{X/S} + 2F \right) \left( p_m + 2F \right). \tag{1}$$



### 7.1. Area of transistor-like devices

First we deal on the same footing with the **area** estimates of transistor-like devices

(**CMOS HP, CMOS LP, HomJTFET, HETJTFET, gnrTFET, BisFET, SpinFET**)

and the area of the **STT/DW**, which all have the same or similar circuit architecture.

The sizes of the inverter with fanout 4, 2 input NAND fanout-1, and 2 input XOR fanout-

1 gates are calculated using the factors in Table 5, which are derived from layout

diagrams. The interconnect length is important for calculating its capacitance and is

estimated as

$$l_{ic} = 5 p_m.$$

( 2 )

This interconnect's area is not an additive term in the area calculation of the circuits,

since it is positioned in the metal-1 layer. However, when interconnect area is needed to

properly route interconnects, we account for the extra empty area needed around the gates

and 1-bit adder cells by introducing the empirically derived area overhead factors listed

in Table 5. These factors work as follows. The area of a gate is obtained by estimating the

gate length and width of the layout cell, and then multiplying it by the gate overhead

factor. We assume that the inverter is made with the standard transistor width $w_X$. We

take the n- and p-transistors to be of the same width due to an assumption of them having

approximately equal on-current. The usual electrode pitch would accommodate

transistors of width $w_X$. Extra width of transistors results in an increase of the gate area.

So for the fanout-4 inverter its area is (Figure 29)



$$a_{inv4} = 2p_m \cdot (3p_m + 2w_X) M_{gate}.$$ ( 3 )

For the NAND2 gate, we make the width of transistors in the pull-down network (nFET in this case) to be twice the standard width. Thus the area of fanout 1 two-input NAND is (Figure 29)

$$a_{nand} = 3p_m \cdot (3p_m + 3w_X) M_{gate}.$$ ( 4 )

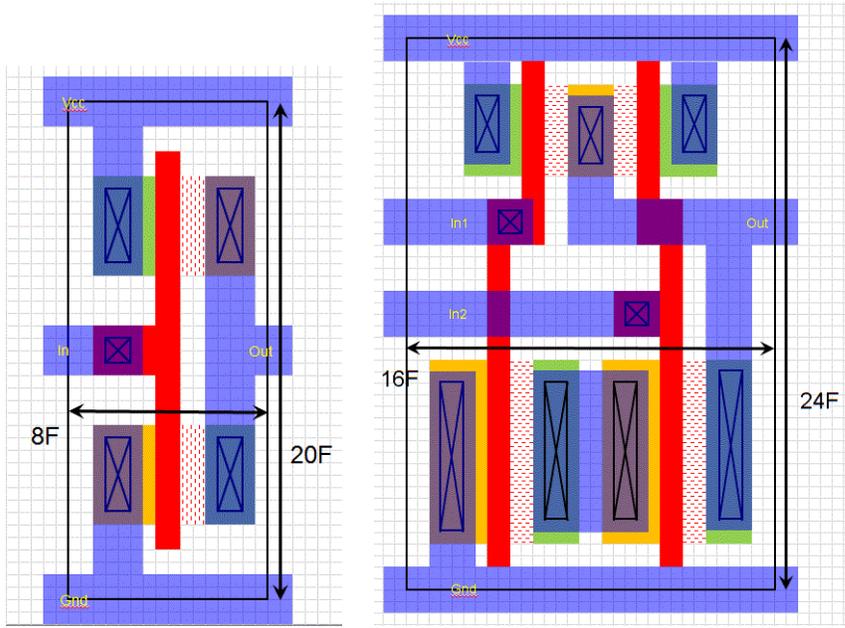

**Figure 30. Layout and area estimation for a TFET inverter and a 2-input NAND gate.**

A tunneling FET needs to have the source and drain to have doping of opposite polarities close to each other. That would violate the traditional scalable CMOS design rules related to p-doped diffusion in an n-well (as we show in Figure 30). We adopt a new design rule of minimum spacing of 2λ between n- and p-diffusion. Thus a new process flow needs to be developed. Also TFETs commonly have gate-drain underlap [11], which we set at the most stringent value of 2λ. [Note that a version of a layout of vertical TFET performed



according scalable MOS design rules was shown in [66]]. With these two changes, the inverter for TFET has the same area as CMOS, while the area of NAND2 is increased to

$$a_{nand,tfet} = 4p_m \cdot \left(3p_m + 3w_X\right)M_{gate} \, . \qquad\qquad (5)$$

In spite of the fact of an addition of a magnetic tunnel junction, the size of the SpinFET cell is the same as that of CMOS, see Figure 31.

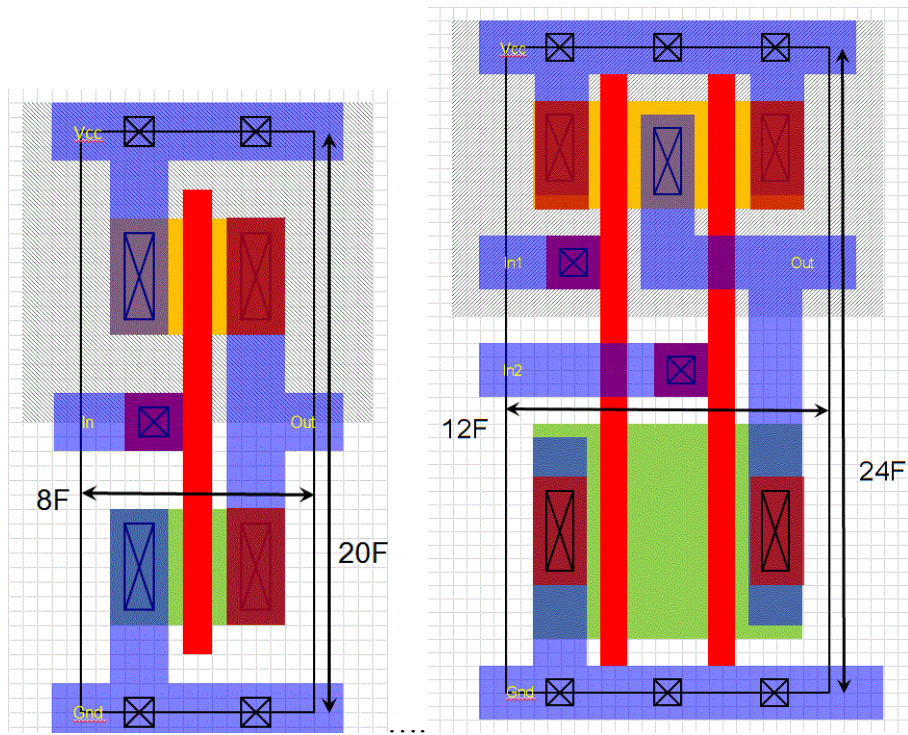

**Figure 31. Layout and area estimation for a SpinFET inverter and a 2-input NAND gate.**

The layouts of BisFET circuits, according to the scheme in Figure 12, are shown in Figure 32. It is apparent that the area for BisFET circuits is larger due to many additional electrodes.



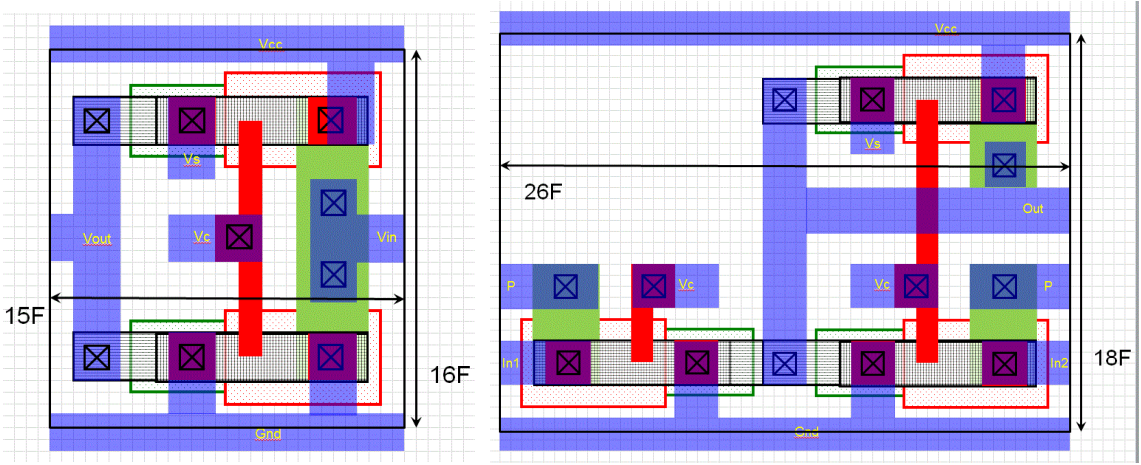

**Figure 32. Layout and area estimation for a BisFET inverter and a 2-input NAND gate.**

The scheme of a 1-bit of a full adder [67] (Figure 33) shows that it consists of 2 XOR-gates with two inputs and 3 AND/OR gates with two inputs, and thus its area is approximately

$$a_1 = \left(2a_{xor} + 2a_{nand}\right) M_{bit}.$$  (6)

We approximately set the XOR area to $a_{xor} = 3a_{nand}$. Note that the circuit diagram of a BisFET-based adder, Figure 34, (despite a different principle of device operation) has in essence the same structure and is therefore treated on the same footing.

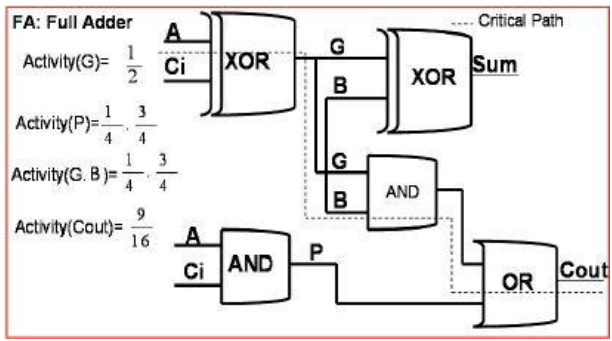

**Figure 33. Transistor-like adder block diagram [67].**



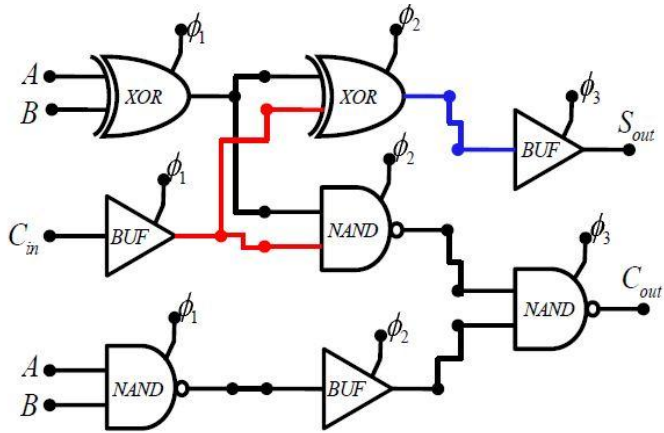

**Figure 34. Bilayer pseudospin (BisFET) adder block diagram [68].**

The area of the full 32 bit adder is calculated as

$$a_{32} = 32a_1 M_{add} \,.$$ ( 7 )

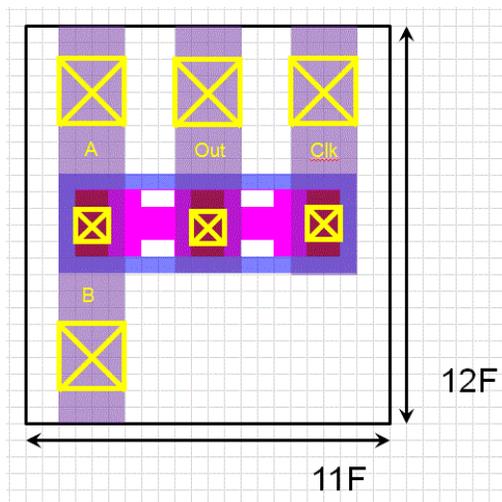

**Figure 35. Layout and area estimation for a STT/DW 2-input AND/OR gate.**

## 7.2. Area of STT/DW

The structure of **STT/DW** is somewhat different and presents a special case. Three

independent electrodes (two inputs and one output) are required for one device which

represents a NAND gate (Figure 15 and Figure 35). Therefore



$$a_{STT} = \left(w_{X/S} + F\right)\left(2p_m + 2F\right). \tag{8}$$

$$a_{nand} = a_{STT}M_{gate}. \tag{9}$$

$$a_{inv4} = 4a_{STT}M_{gate}. \tag{10}$$

The adder consists of "two 3-fan-out NANDs, one 2-fan-out NAND, six 1-fan-out NANDs, and nine COPY elements" [26], see Figure 36. We neglect the area of copy elements and thus estimate the area of a one-bit adder as

$$a_1 = 14a_{nand}M_{bit}. \tag{11}$$

Note that here and on we do not explicitly consider the area occupied by driving transistors for spintronic circuits and just show the contacts to them, like in Figure 35. The reason for that is that spintronic elements are placed between metallization layers, while the transistors are at the semiconductor level. We also assume that the area of spintronic circuits is limited by the size of spintronic elements rather than transistors.

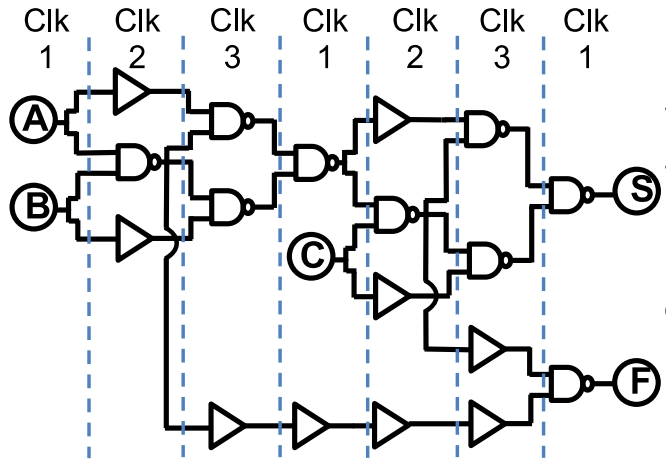

**Figure 36. Spin transfer torque / domain wall (STT/DW) adder block diagram.**



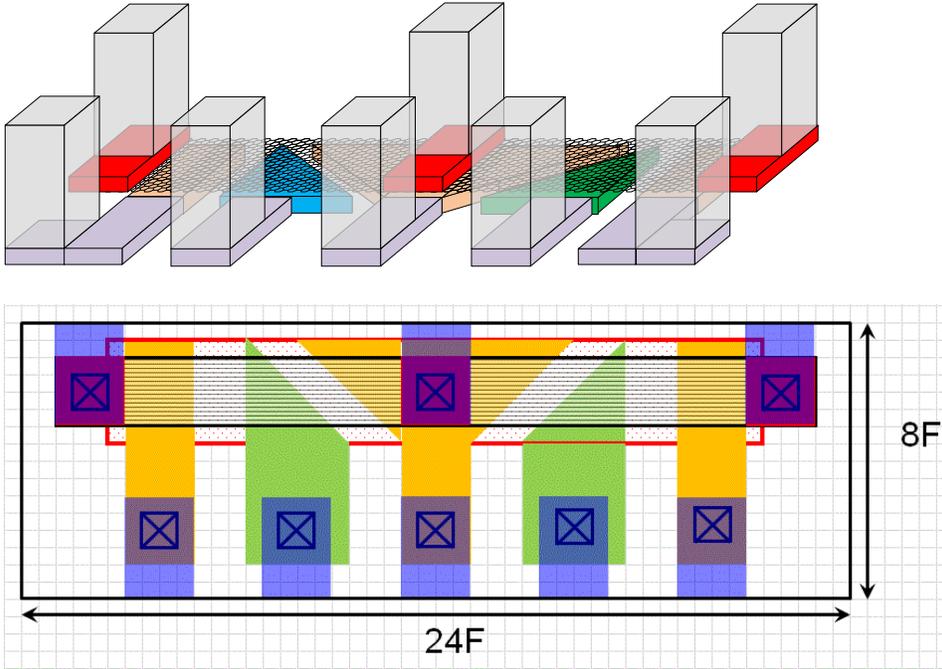

**Figure 37. 3D scheme (top) and layout (bottom) of a GpnJ MUX, after [69].**

### 7.3. Area of GpnJ

The layout of a single MUX is shown in Figure 37 and a GpnJ adder is shown in Figure 38. Red, white and blue rectangles are top electrodes, and yellow and green triangles are p- and n-doped areas of graphene requiring bottom electrodes. One rectangle consisting of 4 graphene triangles realizes a MUX; it contains 5 electrodes along its length and two along its width. We estimate the area of a MUX (not limited by the width of graphene is $w_g = 2F$ )

$$a_{mux} = 2 p_m \cdot 6 p_m .$$

( 12 )

We assume that a fanout-4 and a two-input NAND gate can be composed from just one MUX:



$$a_{inv} = a_{mux} M_{gate} . \qquad ( \ 13 \ )$$

$$a_{nand} = a_{mux} M_{gate} . \qquad ( \ 14 \ )$$

A one bit of a full adder consists of 10 MUXes and its area is:

$$a_1 = 10 a_{nand} M_{bit} . \qquad ( \ 15 \ )$$

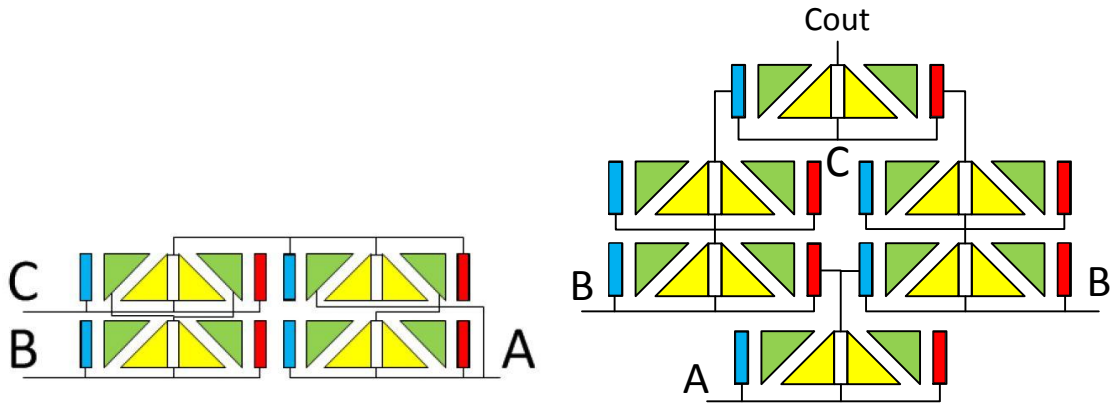

**Figure 38. GpnJ adder scheme. The circuits for generating the sum (left) and the carry (right) signals.**

### 7.4. Area of spintronic devices

Lastly, we treat the **area** of spintronic devices (**SMG, STOlogic, ASLD, SWD, and NML**) similarly, because they are based on majority gates. The example of SMG is used in Figure 39, but we claim that the area estimate is applicable to all spintronic devices. The size of the majority gate is taken as a size of one cross in Figure 39:



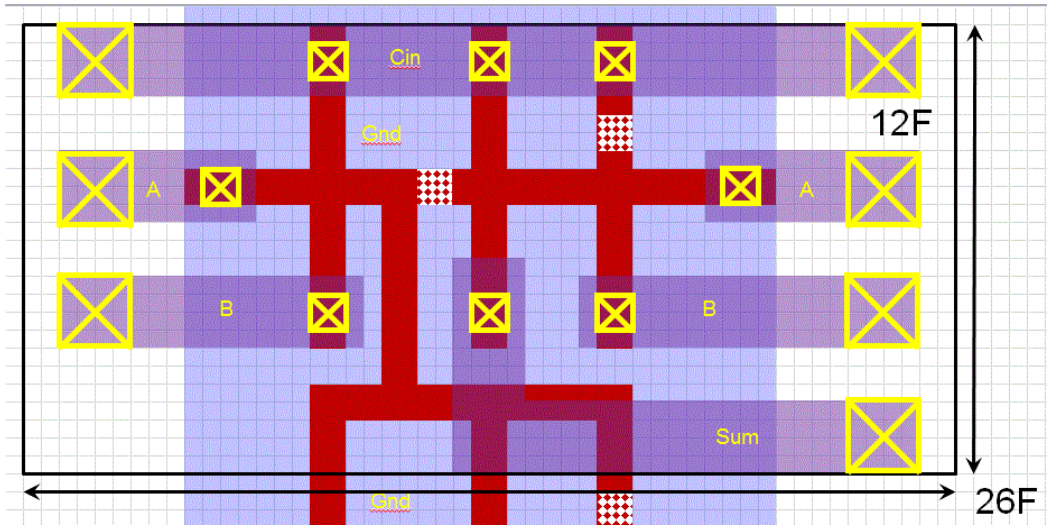

**Figure 39. Layout of a SMG adder composed of three majority gates.**

We approximate the length of the majority gate by

$$l_{maj} = 2 p_m .$$

( 16 )

and the area of a majority gate is

$$a_{maj} = l_{maj}^2 .$$

( 17 )

with the exception of **STOlogic**, where an additional factor of 3 is introduced to account

for wiring of driving transistors in accordance to the layout, see Figure 40.



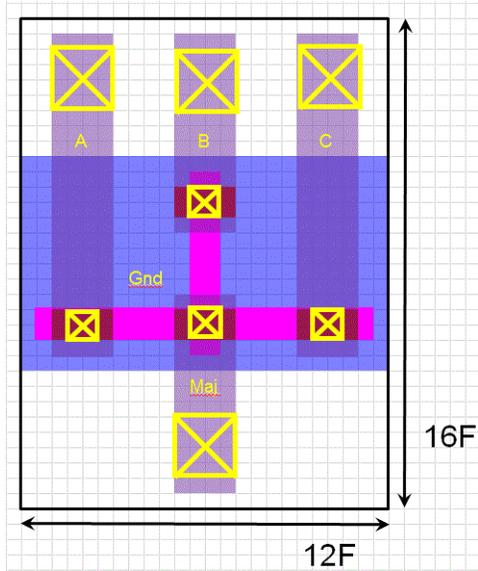

**Figure 40. Layout of a STOlogic majority gate.**

The inverter with fanout of 4 and a NAND gate can be implemented with just one majority gate

$$a_{inv4;nand} = a_{maj} M_{gate}.$$  ( 18 )

The 1-bit full adder's area is determined by the number of majority gates required:

$$a_1 = N_{maj} a_{maj} M_{bit}.$$  ( 19 )

We assume that $N_{maj} = 3$ majority gates are needed to form a one-bit adder, as it has been pointed out in [70]. In some cases fewer majority gates are required or they admit a denser packing. In these cases we adjust the effective number of majority gates to be $N_{maj} = 2$ for **SWD** (see Figure 41) and $N_{maj} = 1$ for **NML** (see Figure 42).



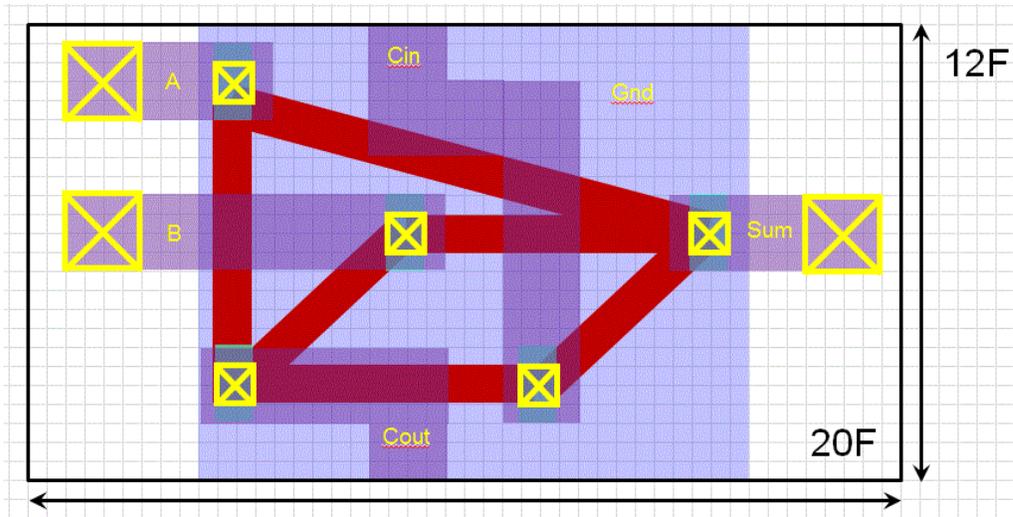

**Figure 41. Layout of a SWD one-bit adder.**

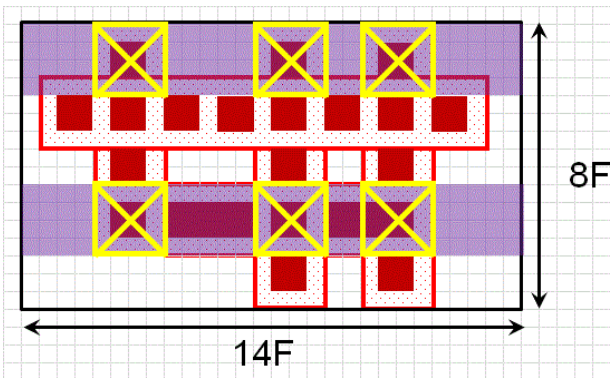

**Figure 42. Layout of a NML one-bit adder.**

While **ASLD** has 2 majority gates per adder due to a clever use of the device's functionality, we still set $N_{maj} = 3$ to correctly account for the cell area (see Figure 43).



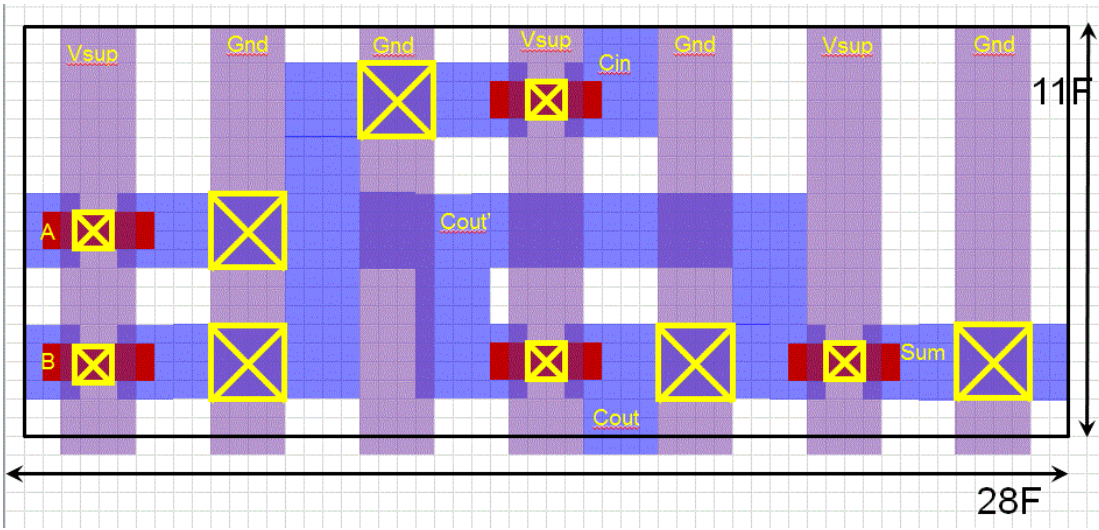

**Figure 43. Layout of ASLD one-bit adder.**

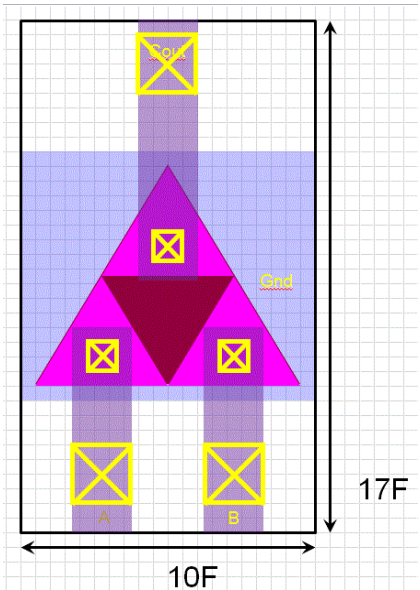

**Figure 44. Layout of a triangular element of the STT triad.**

### 7.5. Area of STT triad

The area of a triangle element in Figure 45 is estimated to be 17x10F according to the layout in Figure 44. The area of a NAND gate is estimated the same way too. An inverter with fanout of four requires four such triangles.



$$a_{inv4} = 4a_{maj}M_{gate}.$$  ( 20 )

A 1-bit full adder requires 9 triangles, therefore

$$a_1 = 9a_{nand}M_{bit}.$$  ( 21 )

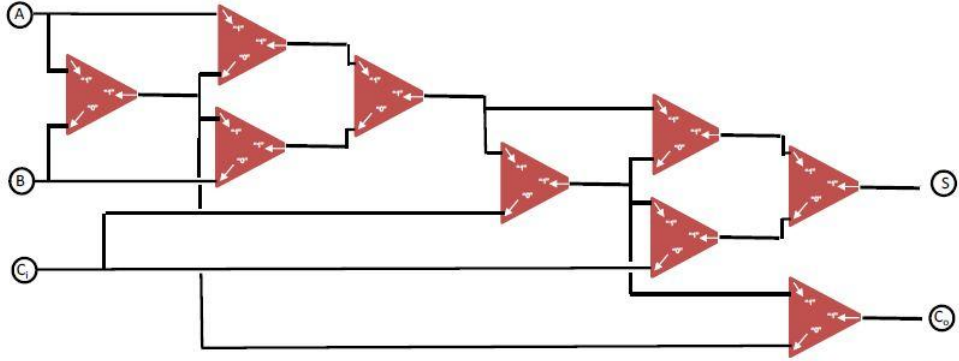

**Figure 45. Spin transfer torque triad (STTtriad) one-bit adder [29].**

The resulting estimates for the areas of beyond-CMOS devices are summarized in Table 6. Note that the intrinsic elements of spintronic logic are larger than those of electronic logic, but the areas of the adder are smaller. This is due to a richer functionality of the spintronic majority gates which enables circuits with fewer elements.

| F*F | Device | INVFO4 | NAND2 | Adder-1b | Adder-32b |
|---|---|---|---|---|---|
| CMOS HP | 36 | 240 | 432 | 5832 | 279940 |
| CMOS LP | 36 | 240 | 432 | 5832 | 279940 |
| HomJTFET | 36 | 240 | 576 | 7776 | 373250 |
| HETJTFET | 36 | 240 | 576 | 7776 | 373250 |
| gnrTFET | 36 | 240 | 576 | 7776 | 373250 |
| GpnJ | 192 | 288 | 288 | 4320 | 207360 |
| BisFET | 24 | 360 | 702 | 9477 | 454900 |
| SpinFET | 36 | 240 | 432 | 5832 | 279940 |
| STT/DW | 20 | 120 | 30 | 630 | 30240 |
| SMG | 64 | 64 | 64 | 288 | 13824 |
| STTtriad | 170 | 1020 | 255 | 3443 | 165240 |
| STOlogic | 128 | 128 | 128 | 576 | 27648 |
| ASLD | 64 | 64 | 64 | 192 | 9216 |
| SWD | 64 | 64 | 64 | 192 | 9216 |
| NML | 64 | 64 | 64 | 96 | 4608 |



**Table 6. Estimate of size of logic circuits, in units of $F^2$.**

## 8. General considerations for switching time and energy

The switching time and energy of the devices strongly depends on the voltage and current at which they run. The most fruitful avenue for improvement of beyond-CMOS devices will be to find a lower operating supply voltage [71] - lower-power alternatives with resulting large improvement in computing efficiency. In this study we use a relatively low supply voltage of 10mV for spin torque switching of spintronic devices. The reason this is possible is that magnetization switching by a current (see Section 11) is unlike transistor switching. Charging the gate capacitor of a transistor results in raising and lowering of a potential barrier, which needs to be several times higher than the thermal energy kT (which corresponds to 26mV) [2] in order to ensure a good turn-off of the transistor and to suppress standby power dissipation. The potential barrier separating the logical states is not changed in the magnetization switching, therefore such a consideration does not limit the supply voltage.

Another factor relevant to both electronic and spintronic circuits is a distribution of power and ground. In order to minimize active power, the voltage controlling power needs to be switched on only when a device is being switched. It may be problematic to switch a power network at a supply voltage smaller than the threshold voltage of a transistor. By setting this low supply voltage for spintronic devices, we implicitly neglect the dissipated power in the low voltage power supply source and in the network for the power and ground distribution. Even though this contribution is significant, a correct account of it goes beyond the scope of this study. Therefore we provide optimistic



estimates for switching delay and energy of the circuits with a voltage of 10mV. For comparison we provide a more realistic calculation with supply voltage of 100mV.

For voltage-controlled switching of spintronic devices we always use the supply voltage of 100mV. It is sufficient, considering that the necessary electric field proves to be small. In this case too, this voltage is not related to raising and lowering the energy barrier and is not connected to standby power dissipation.

Table 7 contains the input parameter values we use for the benchmark calculations for each device. Such values need to be obtained from device-level simulations of the device characteristics. If these inputs change, so might the overall conclusions from the device comparison. For electronic devices the on-current $I_{on}$ is taken for such inputs. For spintronic devices $I_{on}$ designates the current of the driving transistors, and we utilize the high-performance CMOS device for this purpose.

| | Supply voltage $V_{dd}$, V | | $I_{on}$, A/m | $M_{cadj}$ |
|---|---|---|---|---|
| CMOS HP | 0.73 | | 1805 | 1 |
| CMOS LP | 0.3 | | 2 | 1 |
| HomJTFET | 0.2 | | 25 | 0.5 |
| HETJTFET | 0.4 | | 500 | 0.5 |
| gnrTFET | 0.25 | | 130 | 0.5 |
| GpnJ | 0.7 | | 3125.9 | 1 |
| BisFET | 0.6 | | 900 | 2 |
| SpinFET | 0.7 | | 700 | 2 |
| | for current switching | for voltage switching | | |
| STT/DW | 0.01 | N/A | 420 | 1 |
| SMG | 0.01 | 0.1 | 1805 | 1 |
| STTtriad | 0.01 | 0.1 | 1805 | 1 |
| STOlogic | 0.01 | N/A | 2344 | 1 |
| ASLD | 0.01 | N/A | 2344 | 1 |
| SWD | 0.01 | 0.1 | 1805 | 1 |
| NML | 0.01 | 0.1 | 1805 | 1 |

**Table 7. Device parameters: supply voltage, on-current, and capacitance adjustment factor.**



Since most devices (including spintronic) involve charging and discharging of a capacitor of some sort, charging/discharging make a major contribution to their performance. We treat capacitances consistently assuming that the most advanced gate stack is available for all of the devices. The ideal single gate dielectric capacitance per unit area (F/m$^2$) is determined via the equivalent oxide thickness

$$c_{ga} = \frac{\varepsilon \varepsilon_0}{EOT}.$$  ( 22 )

This capacitance, through the use of *EOT*, includes both the dielectric capacitance and the semiconductor (aka quantum) capacitance, while $EOT_0$ would include only the dielectric capacitance. The ideal single gate capacitance per unit length (F/m) is

$$c_{gl} = c_{ga} L_g$$  ( 23 )

The device-dependent adjustment factor $M_{cadj}$ specifies by how much the intrinsic device gate capacitance is larger (or smaller) than the capacitance of a single gate dielectric. We take it larger for BisFET and SpinFET due to the capacitance of additional elements in the device, and we take it smaller for tunneling devices to account for a smaller gate charge in the on-state, see Table 7. We also include an additional factor of 2 for the **BisFET** to account for the top and bottom gates. The expression $M_{cpar}$ indicates the value of the parasitic capacitance (gate-to-source, gate-to-drain (Miller), gate-to-contact, etc.) in terms of the ideal single gate one. We take this factor to be the same for all devices. The total capacitance (F/m) with the parasitics and the device-dependent adjustment factor as well as the contact capacitance $C_c$ is



$$c_{tl} = c_{gl}\left(M_{cadj} + M_{cpar}\right) + C_c \qquad (24)$$

We estimate the capacitance of wires per unit length (F/m) by using the equations [72] for the capacitance of a line surrounded by two ground planes and two neighboring lines.

$$c_{il} = 2\left(c_{gr} + c_{ltol}\right) \qquad (25)$$

We take geometrical parameters as described in Table 5. The capacitance (F, farads) of a typical interconnect between gates is proportional to its length defined in Eq. ( 2 ).

$$C_{ic} = c_{il}l_{ic} . \qquad (26)$$

For F=15nm, these capacitances are $c_{il} = 126aF / \mu m$ and $C_{ic} = 37.8aF$ .

## 9. Switching time and energy for electronic devices

The switching **time and energy** of transistor-like devices (**CMOS HP, CMOS LP, HomJTFET, HETJTFET, gnrTFET, SpinFET, BisFET**) are treated in the same manner. Even though the SpinFET is a spintronic device, the benchmarks considered here use only static combinational logic circuits, in which the magnetization state of SpinFET is not reconfigured.

### 9.1. Intrinsic values for electronic devices

We use extremely simple algebraic equations to estimate their performance. The device capacitance is

$$C_{dev} = c_{il}w_X . \qquad (27)$$

and the device on-current is

$$I_{dev} = I_{on}w_X . \qquad (28)$$

The intrinsic switching time and energy of the device is



$$t_{\text{int}} = C_{dev}V_{dd} / I_{dev}, \quad E_{\text{int}} = C_{dev}V_{dd}^2 . \tag{29}$$

We approximate the time and energy of an interconnect as

$$t_{ic} = 0.7C_{ic}V_{dd} / I_{dev}, \quad E_{ic} = C_{ic}V_{dd}^2 . \tag{30}$$

We also include an additional factor of 2 in the interconnect delay and energy for the

**BisFET** to account for more complex interconnects to the device.

### 9.2. Switching time and energy of GpnJ.

The resistance of the graphene element is composed of the collimation resistance and the

contact resistance

$$R_{dev} = R_{coll} + R_{cont} . \tag{31}$$

Graphene resistivity is estimated according to [73]. The collimation resistance

$$R_{coll} = R_q / M_q , \tag{32}$$

is determined by the number of quantum modes that can propagate in the graphene sheet

$$M_q = 2k_F w_g / \pi , \tag{33}$$

which, in turn, is determined by the Fermi momentum set by applied bias $V$ , which

causes electrostatic p- or n-doping of graphene:

$$k_F = \frac{E_F}{\hbar v_F} , \tag{34}$$

where $E_F$ is the Fermi energy of carriers in graphene [73].

$$E_F = \frac{1}{\gamma EOT_0}\left(\sqrt{\varepsilon^2 + 2\varepsilon\gamma e V_g EOT_0} - \varepsilon\right), \tag{35}$$

where $V_g = V_{dd} / 2$ and the constant



$$\gamma = \frac{1}{\varepsilon_0} \left( \frac{e}{2\pi \hbar v_F} \right)^2.$$ ( 36 )

The contact resistance is inversely proportional to graphene width:

$$R_{cont} = R_c / w_g.$$ ( 37 )

The on-current per length in the device is

$$I_{on} = V_{dd} / (R_{dev} w_g).$$ ( 38 )

The device capacitance $C_{dev}$ which is being switched is that of the middle and side gates (shown in orange in Figure 37). Their total area is *9F* by *3F*. To obtain the total capacitance of the switched MUX one needs to multiply it by $c_{tl}$. The interconnect capacitance $C_{ic}$ are calculated the same way as for other electronic devices, with accounting of the area of the graphene device. The quantum capacitance is included in *EOT*. The intrinsic device delay and the interconnect delay are

$$t_{int} = C_{dev} R_{dev}, \ t_{ic} = C_{ic} R_{dev}$$ ( 39 )

The device intrinsic switching energy $E_{int}$ and the interconnect energy $E_{ic}$ are calculated the same way as for other electronic devices.

| PARAMETER | TYPICAL VALUE |
|---|---|
| Time factor for inverter, $M_{tinv}$ | 0.8 |
| Energy factor for inverter, $M_{Einv}$ | 0.8 |
| Time factor for adder, $M_{tadd}$ | 1.4 |
| Energy factor for adder, $M_{Eadd}$ | 0.3 |
| Parasitic capacitance factor, $M_{cpar}$ | 1.5 |

**Table 8. Circuit performance parameters.**

### 9.3. Circuit values for electronic devices.

The performance of simple circuits is calculated via the empirical factors in Table 8 that are chosen to approximately agree with the simulations done with PETE [67]. They approximately relate to the estimates obtained from comparing the logical efforts of these gates. For the FO4 inverter, the fanout factor of 4 multiplies the expressions for delay and energy:

$$t_{inv} = 4\left(M_{tinv}t_{\text{int}} + t_{ic}\right), \quad E_{inv} = 4\left(M_{Einv}E_{\text{int}} + E_{ic}/2\right). \qquad (40)$$

For the 2-input NAND gate with fanout of 1

$$t_{nand} = M_{tinv}t_{\text{int}} + t_{ic}, \quad E_{nand} = 2M_{Einv}E_{\text{int}} + E_{ic}, \qquad (41)$$

where the factor of 2 in the expression for the energy corresponds to two transistors in the pull-up or pull-down networks. For the XOR gate

$$t_{xor} = 3t_{nan}, \quad E_{xor} = E_{nan}. \qquad (42)$$

For the 1-bit full adder

$$t_1 = M_{tadd}\left(t_{xor} + 2t_{nan}\right), \quad E_1 = M_{Eadd}\left(2E_{xor} + 3E_{nan}\right). \qquad (43)$$

### 9.4. Circuit values for GpnJ

**GpnJ** constitutes a special case. Here an adder contains four MUXes, but the critical path, from carry-in to carry-out, traverses just one MUX. The energy involves switching all 10 MUXes



$$t_1 = M_{tadd}\left(t_{\text{int}} + t_{ic}\right), \quad E_1 = M_{Eadd} 10\left(E_{\text{int}} + E_{ic}\right).$$ ( 44 )

In order to approximately obtain the time and energy benchmarks for the adder, the benchmarks for one bit are multiplied by the number of bits = 32.

$$t_{32} = 32t_1 \ , \quad E_{32} = 32E_1 \ .$$ ( 45 )

## 10. Common methods of magnetization switching

All spintronic devices considered here contain ferromagnets. In order to switch the logical state in them, their magnetization needs to be switched. Presently the preferred way of switching magnetization is with **current-controlled switching**.

One example, where magnetization switching is done with the **magnetic field of current,** is clocking of **NML.**

Magnetization can also be switched by current via the effect of **spin-transfer torque** (STT).

For the case of **in-plane magnetization with STT** it is assumed that the nanomagnet has the aspect ratio of 2 and thus the area and volume of

$$a_{nm} = 2w_S^2, \quad v_{nm} = a_{nm}d_{fm}.$$ ( 46 )

The energy barrier height [74] is determined via the difference of the element of the demagnetization tensor in the plane of the nanomagnet $\Delta N = N_{yy} - N_{xx}$ (which are a function of the ratios of the length, width and thickness and of the shape of a nanomagnet) as follows



$$U_b = \Delta N \mu_0 M_s^2 v_{nm} / 2 .$$ ( 47 )

See parameter definitions in Table 3. We take values approximately corresponding to the

alloy CoFe. The critical current density (after [74]) is

$$J_c = \frac{e \alpha \mu_0 M_s^2 d_{fm}}{\hbar P} .$$ ( 48 )

We chose to operate with the switching current

$$I_{dev} = 3 I_c = 3 J_c a_{nm} .$$ ( 49 )

Then the switching time is described by [74,75]

$$t_{stt} = \frac{e M_s v_{nm}}{g \mu_B P \left( I_{dev} - I_c \right)} \log \left( \frac{2 \pi \sqrt{2 k_B T}}{\sqrt{U_b}} \right) .$$

( 50 )

Thus the energy needed to switch is

$$E_{stt} = I_{dev} V_{dd} t_{stt} .$$

( 51 )

For the case of the **perpendicular magnetization STT**, it is assumed the aspect ratio is

1, i.e.,

$$a_{nm} = w_S^2 , \quad v_{nm} = a_{nm} d_{fm} .$$ ( 52 )

The barrier height is determined by the perpendicular magnetic anisotropy (PMA) [76]

$$U_b = K_u v_{nm} .$$ ( 53 )

We take parameters corresponding to a material with a relatively small saturation

magnetization and high PMA such as CoPtCrB [77]. The critical current density [76] is

$$J_c = \frac{2 e \alpha K_u d_{fm}}{\hbar P} .$$ ( 54 )

The switching time is given by the same expression, Eq. ( 50 ).



A more energy efficient, but less technologically mature means of switching of magnetization is the **voltage-controlled switching**.

Voltage-controlled switching with an adjacent **multiferroic material** occurs due to the effective magnetic field at the interface, that arises through the exchange bias effect. For this case, we take parameters corresponding to bismuth-ferrite (BiFeO3, BFO) [59,78]. To switch the ferroelectric polarization of the multiferroic material, the capacitance must be charged from a power supply with a CMOS transistor (we will use the parameters for CMOS HP here). The switching occurs in a hysteretic manner and is characterized by the critical field $E_{mf}$ and remanent polarization $P_{mf}$. See parameters values in Table 3. The total charge that needs to be supplied is

$$Q_{mf} = P_{mf} w_S^2 + c_{tt} w_X V_{dd}.$$

(55)

The required thickness of the multiferroic is

$$d_{mf} = \frac{V_{dd}}{E_{mf}}.$$ ( 56 )

The charging energy is

$$E_{mf} = Q_{mf} V_{dd}.$$ ( 57 )

The charging time is

$$t_{mf} = Q_{mf} / I_{dev}.$$ ( 58 )

and the magnetization switching time is

$$t_{mag} = \frac{\pi}{2\gamma B_{me}}.$$

(59)

The total switching time is the combination of the two.



Another way of doing voltage-controlled switching is with an adjacent piezoelectric material. Changing the polarization in a piezoelectric material causes strain, and the stress at the interface switches magnetization in the ferromagnet by **the magnetostrictive effect**. In an example of switching using a the piezoelectric material such as lead magnesium niobate-lead titanate (PMN-PT) [60], the switching also has a hysteretic character. The expressions are similar to Eqs. (55)-(59) with subscript "ms", except the polarization is determined via the dielectric constant

$$P_{ms} = \varepsilon_0 \varepsilon_{ms} \mathrm{E}_{ms}.$$  ( 60 )

We also describe here for completeness a similar magnetostrictive switching by the **linear magnetoelectric effect** [62]. The strength of coupling is expressed by the magnetoelectric coefficient (Table 3). However none of the devices in the present study envision using this type of switching. The electric field in the piezoelectric is

$$\mathrm{E}_{me} = \frac{V_{dd}}{d_{me}}.$$  ( 61 )

The resulting magnetic field [62] is

$$B_{me} = \alpha_{me} \mathrm{E}_{me}.$$  ( 62 )

The induced polarization is

$$P_{me} = \varepsilon_0 \varepsilon_{me} \mathrm{E}_{me}.$$  ( 63 )

Another way of switching is by means of a voltage change of surface anisotropy [79]. We do not cover this method in this paper.

The estimates in this section are used in benchmarking multiple spintronic devices. Note that we incorporate only the energy contribution of driving transistors, but not reading circuits (e.g. sense amplifiers). In some cases that latter contribution can become significant.



# 11. Switching time and energy for spintronic devices

In Section 10 we calculated the switching time and energy for an intrinsic device switching magnetization. Now we expand it to calculating the performance of spintronic majority gate circuits. Table 9 specifies which method of controlled switching of magnetization is assumed, e.g. current-controlled or voltage-controlled, and therefore which estimates from Section 10 are used for the device's intrinsic switching time and energy. For a few of the devices, their method of magnetization switching is specific, and is described below. As the reader can see, the magnetostrictive rather than multiferroic option for voltage-controlled switching was selected for the benchmarking since it gives slightly better projections. For some of the devices we do not envision (indicated with N/A in Table 9) a voltage-controlled option using the principle of the device operation relying exclusively on spin transfer torque.

|          | Current-control    | Voltage-control  |
|----------|--------------------|------------------|
| STT/DW   | domain wall STT    | N/A              |
| SMG      | perpendicular STT  | magnetostrictive |
| STTtriad | in-plane STT       | magnetostrictive |
| STOlogic | perpendicular STT  | N/A              |
| ASLD     | perpendicular STT  | N/A              |
| SWD      | RF antenna         | magnetostrictive |
| NML      | field of current   | magnetostrictive |

**Table 9. Options for current- and voltage controlled switching of spintronic devices.** Here we describe device-specific estimates of the switching time and energy. SMG is using the intrinsic values from Section 10 directly.

## 11.1. ASLD intrinsic values

The net electric current flows from the top electrode to the ground and produces the spin polarization in the interconnect between nanomagnets. This current determines the Joule



heating dissipation. The spin polarized electrons diffusion in both directions between two nanomagnets which causes switching of nanomagnets by spin torque. We introduce an additional factor of 1.5 to account for the spin-polarized diffusion current being smaller than the electric current.

### 11.2. STT-DW intrinsic and circuit values

The magnetization is switched by spin transfer torque of electrons crossing a domain wall and thus causing its motion. The required current per unit length is (see Table 3 and Table 4 for material and device parameters)

$$I_{on} = 1.5 J_{dw} d_{fm}, \qquad (64)$$

where the factor 1.5 is introduced for the excess of the switching current over the critical one, and this current is passed through the magnetic wire of width $F$. The speed constant corresponding to spin torque in domain walls is [80]

$$v_{stt} = \mu_B P J_{dw} / (e M_s). \qquad (65)$$

The speed of domain walls varies and can be approximated by [81]

$$v_{dw} = 3 v_{stt}. \qquad (66)$$

The intrinsic device (a ferromagnetic wire with domain wall driven through it) switching time will be the time to move the domain wall past the magnetic tunnel junction (length $2F$). The switching energy will be the energy needed to be supplied at the clock such that the domain wall receives sufficient current to switch.

$$t_{int} = 2F / v_{dw}, \quad E_{int} = I_{on} w_S V_{dd} t_{int}. \qquad (67)$$

**STT/DW** is the only spintronic device envisioned with the NAND (rather than majority gate) logic functionality. The interconnects there are the usual electrical ones. Their delay and energy proves to be small compared to intrinsic device values.



For the FO4 inverter

$$t_{inv} = t_{\text{int}} + t_{ic}, \quad E_{inv} = 4\left(E_{\text{int}} + E_{ic}\right).$$ ( 68 )

For the 2-input NAND gate with fanout of 1

$$t_{nand} = t_{\text{int}} + t_{ic}, \quad E_{nand} = E_{\text{int}} + E_{ic}.$$ ( 69 )

For the 1-bit full adder, the critical path goes through two NAND gates and the total energy is proportional to the area of all the NANDs in the adder:

$$t_1 = 2t_{nan}, \quad E_1 = 14E_{nan}.$$ ( 70 )

### 11.3. SWD intrinsic values

We assume that the electrical signal exiting the magnetoelectric cell is a harmonic wave with a frequency $f_{sw} = 100GHz$, the corresponding wavelength of spin waves is $\lambda_{sw} = 30nm$, the resulting speed of spin waves is

$$v_{sw} = f_{sw}\lambda_{sw}.$$ ( 71 )

Then the intrinsic switching time is estimated as

$$t_{\text{int}} = 10 / f_{sw}.$$ ( 72 )

and the interconnect delay is

$$t_{ic} = 2l_{maj} / v_{sw}.$$

( 73 )

### 11.4. STO logic intrinsic values

The estimates for the device parameters follows the perpendicular STT series of equations in Section 10, with the exception that we take a typical frequency of oscillations to be $f_{osc} = 30GHz$, and assume that it has a switching time of



$$t_{\text{int}} = 30 / f_{osc}. \tag{74}$$

The interconnection between spin torque oscillators is done via spin waves. The interconnect delay is calculated via the same expression ( 73 ).

In case of **voltage-controlled switching**, the energy estimates are obtained as in Section 10. In case of **the current-controlled switching**, the spin wave is excited by a magnetic field generated by a current in a wire. The estimate of the switching energy is based on the current required to produce $B_{wi}$ of the magnetic induction at a distance of one pitch of metal-1

$$I_{dev} = 2\pi B_{wi} p_m / \left( \mu \mu_0 \right). \tag{75}$$

We use parameters from Table 3 here. The permeability of the substance surrounding the wire has been demonstrated to have values of 2 to 6, see [82]. Then the dissipated energy in a transmission line with impedance of $Z = 50\Omega$ is

$$E_{\text{int}} = I_{dev}^2 Z t_{\text{int}}. \tag{76}$$

### 11.5. NML intrinsic values

For **NML** [83], the switching time of one nanomagnet in the chain forming majority gates is taken as an empirical value $t_{nm} = 0.1ns$ [84,85,86], which presumably does not change much with the size. The number of nanomagnets (spaced at 2F center to center) across a majority gate is

$$N_{nm} = l_{maj} / \left( 2F \right). \tag{77}$$

We also assume that the interconnect between the majority gates requires one half the number of nanomagnets as the path through a majority gate. Then the intrinsic switching time and the interconnect delay are



$$t_{\text{int}} = t_{nm} N_{nm}.$$ ( 78 )

In case of voltage-controlled switching, the energy estimates are obtained as in Section 10. In case of the current-controlled switching, the magnetic field from a set of electrical wires is applied to clock the nanomagnets, see Figure 46. We take the wire width to be $p_m$, the wire pitch to be $2p_m$, and the aspect ratio of wires $AR = 3$ (corresponding to tall wires). The required current in a wire is

$$I_{dev} = 2B_{wi} p_m / \left( \mu \mu_0 \right).$$ ( 79 )

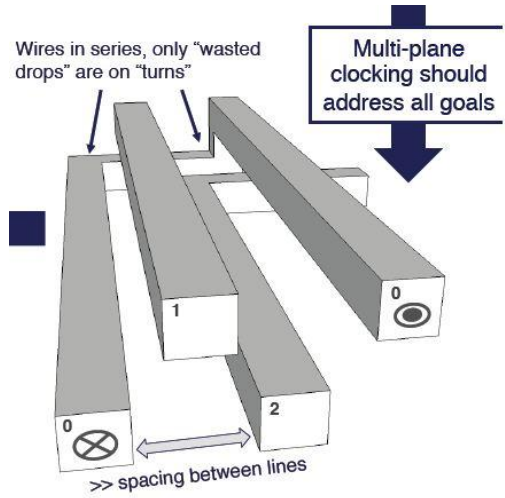

**Figure 46. Electrical wires for clocking NML [87].**

The resistance of the clocking set of wires covering the area is proportional to the area. Each set of wires is used to clock multiple gates simultaneously. To calculate the power dissipation of one majority gate, we substitute its area to the expression for resistance

$$R_{clk} = \frac{\rho a_{maj}}{p_m^3 \cdot AR}.$$ ( 80 )

The energy required to switch over the time of both the device and the interconnect delay



$$E_{\text{int}} = I_{dev}^2 R_{coil} \left( t_{\text{int}} + t_{ic} \right).$$ ( 81 )

For the case of voltage-controlled clocking of NML, we assume that the magnetoelectric (multiferroic or a magnetostrictive material) is driving the area of all nanomagnets in the majority gate

$$a_{NML} = 2N_{nm}2F^2.$$ (82)

and this area is replaces $w_S^2$ in the expression for the energy (55).

### 11.6. Common analysis of spintronic circuits

After the intrinsic device switching time and energy are calculated, the treatment is similar for spintronic devices **SMG, ASLD, SWD, NML, and STO logic**.

We assume that interconnects between gates take the same time to switch as connections within the gate, but they are driven by the energy in the gate (i.e. no additional energy per interconnect). It is only true for short enough interconnect between gates as specified by Eq. ( 2 ). It is applicable unless we explicitly state otherwise for a device.

$$t_{ic} = t_{\text{int}}, \quad E_{ic} = 0.$$ ( 83 )

One exception is **NML**, where the interconnect requires ½ of the time and energy as the intrinsic device (due to smaller number of nanomagnets).

$$t_{ic,NML} = t_{\text{int},NML}/2 \quad E_{ic,NML} = E_{\text{int},NML}/2,$$ ( 84 )

An inverter can be accomplished with one magnetization switching terminal, but the energy is proportional to the number of outputs in a fanout of 4.



$$t_{inv} = t_{\text{int}} + t_{ic}, \quad E_{inv} = 4\left(E_{\text{int}} + E_{ic}\right).$$ ( 85 )

A NAND gate is implemented by a majority gate with three inputs. We assume that it takes the same time to switch as a single input, but the energy is proportional to the number of inputs

$$t_{nand} = t_{\text{int}} + t_{ic}, \quad E_{nand} = 3\left(E_{\text{int}} + E_{ic}\right),$$ ( 86 )

except for **NML**, where the factor of 3 is absent in the last equation because the energy is spent on clocking the whole device rather than separate inputs.

A 1-bit full adder performance is given by the number of required majority gates:

$$t_1 = \min(2, N_{maj})t_{nand}, \quad E_1 = N_{maj}E_{nand} + E_{ic}.$$ ( 87 )

The contributions of interconnects is accounted in majority gates, except that a full adder requires an additional interconnect.

### 11.7. STT triad circuit values

The switching time and energy of **STT triad** is a special case.

We start with the estimates for the switching time and energy in a triangular in-plane nanomagnet, Eqs. ( 50 )-( 51 ), of side length 4.5*F*. The interconnect energy and delay are calculated like any other electrical interconnect, Section 9. The gate switching time and energy contain a factor of 2 for the necessity to reset the triangles before switching. In addition, the switching energy contains a factor of 2 for the two inputs.

$$t_{nand} = 2\left(t_{\text{int}} + t_{ic}\right), \quad E_{nand} = 4\left(E_{\text{int}} + E_{ic}\right).$$ ( 88 )

Additionally, a fanout-four inverter requires a triangle per output with the corresponding factor in the switching energy.



$$t_{inv} = 2\left(t_{\text{int}} + t_{ic}\right), \quad E_{inv} = 16\left(E_{\text{int}} + E_{ic}\right). \tag{89}$$

A 1-bit full adder contains totally 9 triangles and 2 output interconnects (for the sum and the carry) that contribute to the energy. The delay is defined by the critical path through 6 triangles.

$$t_1 = 6t_{nand} + t_{ic}, \quad E_1 = 9E_{nand} + 2E_{ic}. \tag{90}$$

## 12. Comparison of devices

First we assemble summary of intermediate results - the intrinsic device and interconnect times and energies for devices (as defined in Sections 9 and 11). The estimates (Table 10) are performed assuming the use of magnetostrictive switching for spintronic devices which admit it, and spin torque switching for devices exclusively based on it. Zero switching energy of interconnects means that either this energy is neglected compared with the intrinsic device switching energy, or the energy to drive an interconnect is accounted for in the the intrinsic device switching energy.

| device name | Delay, int | Energy, int | Delay, ic | Energy, ic |
|---|---|---|---|---|
| units | ps | aJ | ps | aJ |
| CMOS HP | 0.25 | 19.63 | 0.18 | 20.16 |
| CMOS LP | 92.08 | 3.32 | 66.20 | 3.40 |
| HomJTFET | 3.27 | 0.98 | 3.53 | 1.51 |
| HetJTFET | 0.33 | 3.93 | 0.35 | 6.05 |
| gnrTFET | 0.79 | 1.53 | 0.85 | 2.36 |
| GpnJ | 2.17 | 142.77 | 0.28 | 18.54 |
| BisFET | 1.36 | 22.10 | 1.18 | 27.24 |
| SpinFET | 1.02 | 30.08 | 0.44 | 18.54 |
| STT/DW | 1762.90 | 111.06 | 0.02 | 0.00 |
| STMG | 297.61 | 1.38 | 297.61 | 0.00 |
| STTtriad | 298.03 | 10.92 | 0.02 | 0.38 |
| STOlogic | 1000.00 | 351.60 | 80.00 | 0.00 |
| ASLD | 205.16 | 108.20 | 205.16 | 0.00 |
| SWD | 297.61 | 1.38 | 80.00 | 0.00 |
| NML | 400.00 | 19.31 | 200.00 | 9.65 |

Table 10. Switching delay and switching energy for intrinsic components and interconnects of devices, for magnetostrictive switching.

Now we summarize the results for all circuits composed of the NRI devices under consideration. We plot the switching energy vs. switching delay of the three circuits – fanout-4 inverter, 2-input NAND (see Appendix A), and a 32-bit adder. The plots are done for various cases ofmagnetization switching.

We point out the extremely important role of voltage-controlled magnetization switching as opposed to current-controlled switching. With current controlled switching (Figure 47 - Figure 48) spintronics circuits prove to switch significantly slower and require higher switching energy than electronic circuits. This is related to the limitations of spin torque switching. The energy could be lowered, if operation with a lower supply voltage, 10mV, were possible, but then it may face limitations of critical current fluctuations. For a higher supply voltage, 100mV, the switching energy is correspondingly 10 times higher.



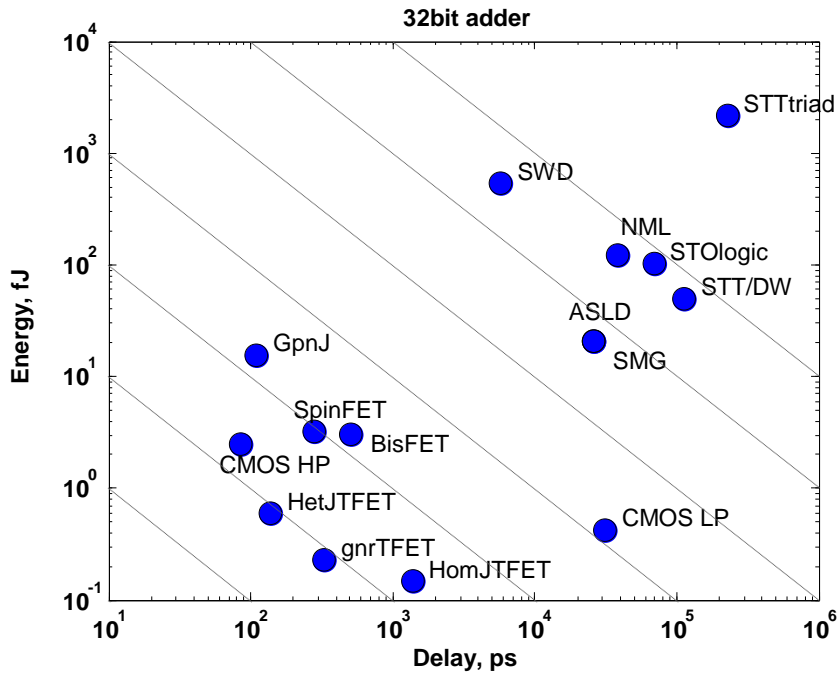

**Figure 47. Energy vs. delay of 32-bit adders. Spintronic devices use current-controlled switching with Vdd=0.01V. The preferred corner is bottom left.**

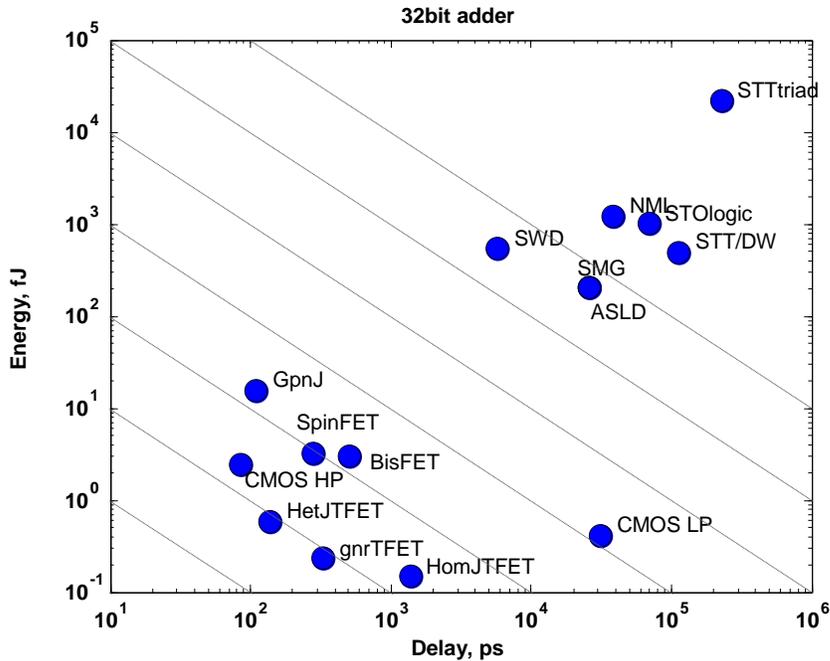

**Figure 48. Energy vs. delay of 32-bit adders. Spintronic devices use current-controlled switching with Vdd=0.1V. The preferred corner is bottom left.**



With voltage controlled switching (Figure 49 - Figure 50) the intrinsic speed and energy of spintronic devices can be significantly improved, and spintronic circuits become competitive with electronic ones. (The exceptions are ASLD, STT/DW and STO logic which inherently rely on spin torque for operation). Magnetostrictive switching results in somewhat better switching energy.

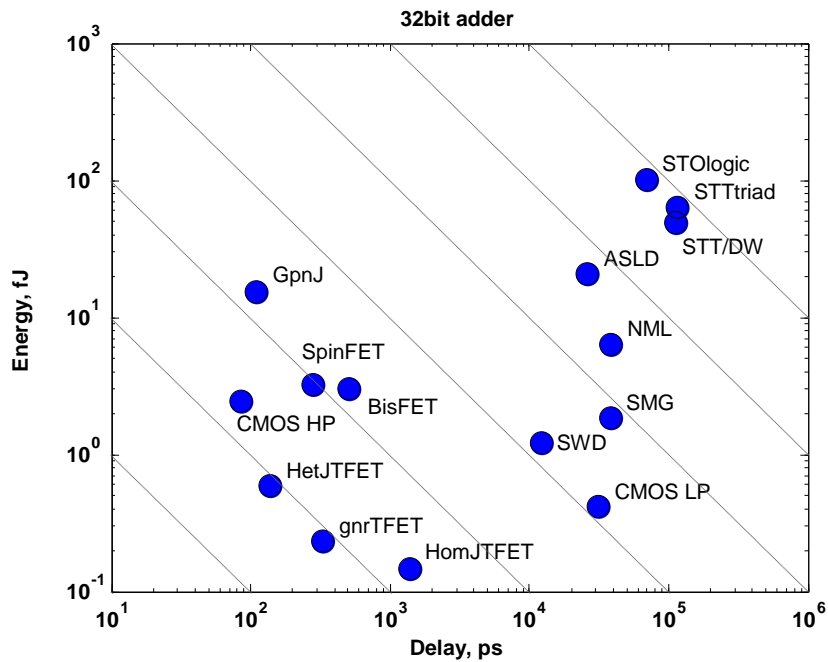

**Figure 49. Energy vs. delay of 32-bit adders. Spintronic devices use multiferroic voltage controlled switching. The preferred corner is bottom left.**



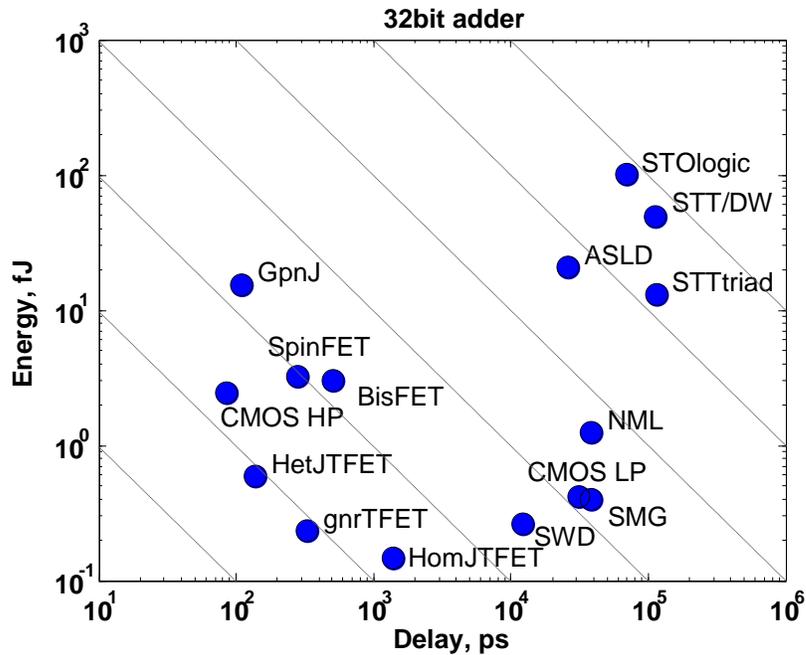

**Figure 50. Energy vs. delay of 32-bit adders. Spintronic devices use magnetostrictive voltage controlled switching. The preferred corner is bottom left.**

Finally we compare (Figure 51) the benchmark results obtained in the present study with the ones reported at the 2011 NRI benchmarking workshop [88]. Some of the devices' benchmarks differ significantly. In some cases (e.g. BisFET) the reason is the different assumptions about devices parameters (on current, voltage). In the case of STT/DW, the difference is due to a dramatic change in the use for logic circuits and also perpendicular magnetic anisotropy materials were used.



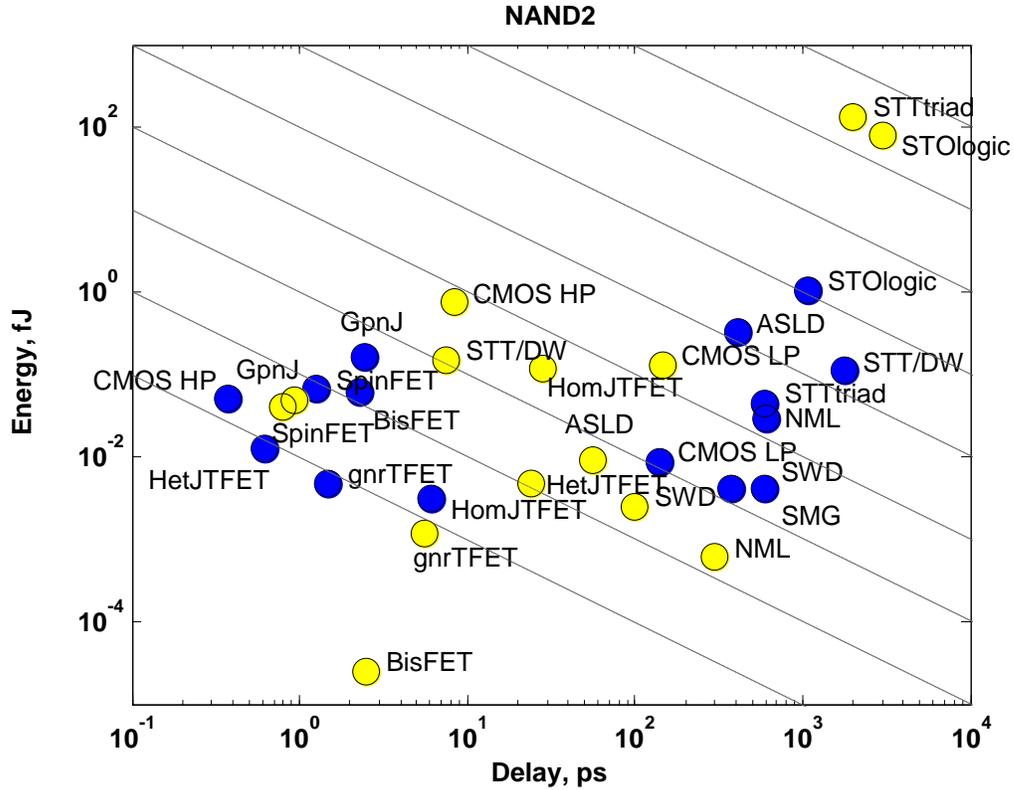

**Figure 51. Energy vs. delay of 2-input NAND gates. Comparison of: NRI benchmark results [5] (yellow dots) with the present study results using magnetostrictive voltage controlled switching (blue dots). The preferred corner is bottom left.**

## 13. Computational throughput and power dissipation

The computational throughput is a measure of useful work performed by a circuit and is defined as a number of integer operations (32 bit additions in this case) per second per unit area. We estimate it as

$$T_{add} = \frac{1}{a_{32}t_{32}}.$$

( 91 )

In the process of computation, the expended energy is dissipated as heat power



$$P_{diss} = T_{add} E_{32}.$$ ( 92 )

Since leakage directly measures stand-by power, it is on its own a metric. Those devices with significant leakage will need power supply gating and this will add to their power. In this calculation we only account for the active power and for the moment neglect the stand-by power (e.g. from the leakage currents). The activity factor is determined by the logic function of the ripple-carry adder and is thus equal to 1/32. We do not introduce additional activity factors which may be pertinent to specific usage models of circuits. We also do not incorporate any pipelining in this calculation, even though some logic technologies are not constrained by power dissipation may produce higher computational throughput by pipelining. For the sake of simplicity we do not consider the use of multi-phase clocks.

As we will see from Figure 49 some of the beyond-CMOS devices have exceptionally low power dissipation and thus are useful for mobile computing. Benchmark numbers for other devices result in a much higher dissipation. Meanwhile there is a practical limit to power dissipation set by the ability of a heat sink to remove power. In this study we (somewhat arbitrarily) set the power density limit ("cap") to $P_{cap} = 10W / cm^2$. Note that it is not the whole power dissipated on chip, since we have not included the contribution of i) long interconnects (longer than bit-to-bit), ii) the clock distribution circuits and iii) other systems on chip. If the dissipated power exceeds the limit, one needs to run the circuits slower or spread them apart, thereby decreasing useful throughput.

Therefore we postulate as a figure merit the computational throughput with a limit on (capped) power and area (fixed);



$$T_{cap} = T_{add} \min(1, P_{cap} / P_{diss}) . \qquad (93)$$

The relationship between capped throughput and dissipated power is shown in Figure 52. The unit of throughput there is peta-integer-operations per second per square centimeter. We see that only the heterojunction TFET is expected to provide higher throughput than high-performance CMOS (and do it with lower dissipated power). BisFET and SpinFET have lower but comparable throughput at the same power. Other tunneling FETs as well as a few of the spintronics devices provide comparable throughput at significantly lower power.

We remind the reader that spintronic logic has the major attribute (advantage) of non-volatility and reconfigurability.

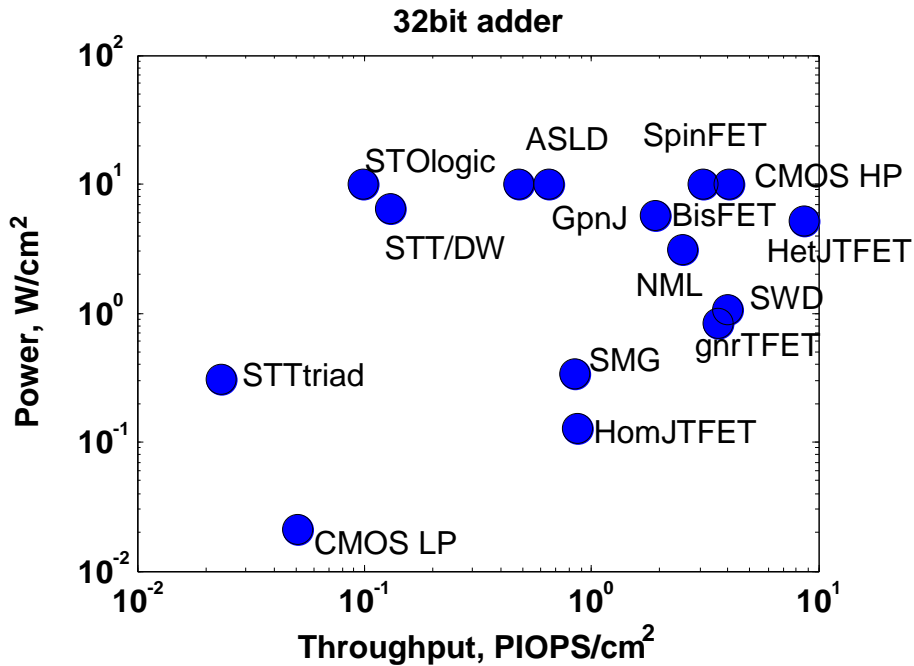

**Figure 52. Throughput vs. dissipated power of devices. The preferred corner is bottom right.**

Alternatively these data are also summarized in the following Table 11.

| device name | Area | Delay | Energy | Power | Throughput | Thr@<10W/cm$^2$ |
|---|---|---|---|---|---|---|



| units | μm² | ps | fJ | W/ cm² | Pops/s/ cm² | Pops/s/ cm² |
|---|---|---|---|---|---|---|
| CMOS HP | 63.0 | 84 | 2.48 | 46.5 | 18.79 | 4.04 |
| CMOS LP | 63.0 | 31331 | 0.42 | 0.0 | 0.05 | 0.05 |
| HomJTFET | 84.0 | 1378 | 0.15 | 0.1 | 0.86 | 0.86 |
| HetJTFET | 84.0 | 138 | 0.59 | 5.1 | 8.64 | 8.64 |
| gnrTFET | 84.0 | 331 | 0.23 | 0.8 | 3.60 | 3.60 |
| GpnJ | 46.7 | 110 | 15.49 | 301.5 | 19.47 | 0.65 |
| BisFET | 102.4 | 508 | 3.00 | 5.8 | 1.92 | 1.92 |
| SpinFET | 63.0 | 282 | 3.20 | 18.0 | 5.63 | 3.13 |
| STT/DW | 6.8 | 112820 | 50 | 6.5 | 0.13 | 0.13 |
| STMG | 3.1 | 38094 | 0.40 | 0.3 | 0.84 | 0.84 |
| STTtriad | 37.2 | 114450 | 13.04 | 0.3 | 0.02 | 0.02 |
| STOlogic | 6.2 | 69120 | 101 | 23.5 | 0.23 | 0.10 |
| ASLD | 2.1 | 26261 | 20.78 | 38.2 | 1.84 | 0.48 |
| SWD | 2.1 | 12084 | 0.26 | 1.1 | 3.99 | 3.99 |
| NML | 1.0 | 38400 | 1.24 | 3.1 | 2.51 | 2.51 |

**Table 11. Summary of comparison of devices for magnetostrictive voltage controlled switching.**

# 14. Discussion of benchmarks

We observe that two major conclusions are derived from this study.

1) For electronic devices, the most promising avenue of improvement is decreasing the operational voltage. Tunneling FETs seem to be the leading option.

2) Spintronic devices have an advantage in implementing complex logic functions with a smaller number of devices/elements. The key factor toward making them competitive with CMOS is using voltage controlled switching.

We would like to stress two additional advantages of spintronic devices that are not captured by the current set of benchmarks – non-volatility and reconfigurability. In order to utilize these advantages new types of circuits need to be designed.



To make the task tractable, we considered only a small set of circuits. In order to cover all elements necessary for an arithmetic logic unit (ALU), one needs to extend this treatment to state elements, latches, multiplexors/demultiplexors, etc.

## 15. Conclusions

An approach with simple but general estimates of benchmark parameters has been described that is applicable to multiple devices and provides a consistent and reproducible methodology that can be used in the benchmarking of NRI devices. By publishing this paper the authors solicit inputs and suggestions from the researchers in the field in order to find uniform consensus on the way to do benchmarking for beyond-CMOS devices.

## 16. Acknowledgements


The authors would like to thank George Bourianoff, Brian Doyle, Charles Kuo, Uygar Avci, Raseong Kim, and Sasikanth Manipatruni for fruitful discussions on beyond-CMOS devices; Arijit Raychowdhury and Charles Augustine for help with the PETE simulator; Kerry Bernstein and Jeff Welser for helpful comments. We are immensely grateful to researchers in Nanoelectronics Research Initiative – Jim Allen, Marc Baldo, Behtash Behin-Aein, Gyorgy Csaba, Jean Anne Currivan, Suman Datta, Supriyo Datta, Kos Galatsis, Yunfei Gao, Sharon Hu, Alexander Khitun, Alexander Kozhanov, Ilya Krivorotov, Ji Ung Lee, Mark Lundstrom, Allan MacDonald, Dejan Markovic, Azad Naeemi, Vijay Narayanan, Michael Niemier, Chenyun Pan, Wolfgang Porod, Kaushik Roy, L. Frank Register, Caroline Ross, Sayeef Salahuddin, Vinay Saripalli, Angik Sarkar,




Alan Seabaugh, Srikant Srinivasan – for providing us a wealth details of their approach to benchmarking beyond-CMOS devices and spending significant time and effort in helping us understand their operation.

## 17. Appendix A.

This section contains the comparison plots for devices in addition to ones in the main text. These provide more perspective on the relative differences amongst the devices.

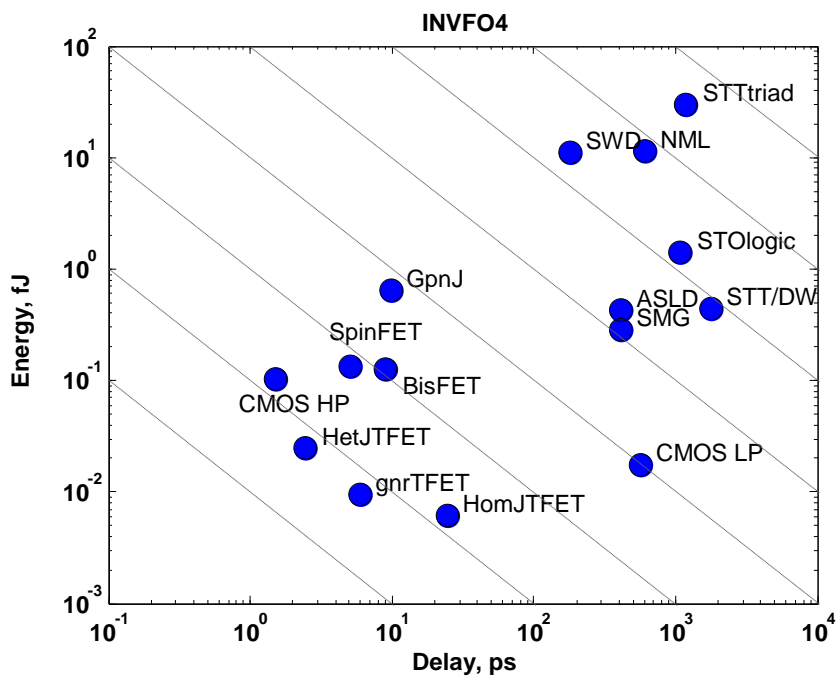

Figure 53. Energy vs. delay of inverters with fanout of 4. Spintronic devices use current-controlled switching with Vdd=0.01V. The preferred corner is bottom left.



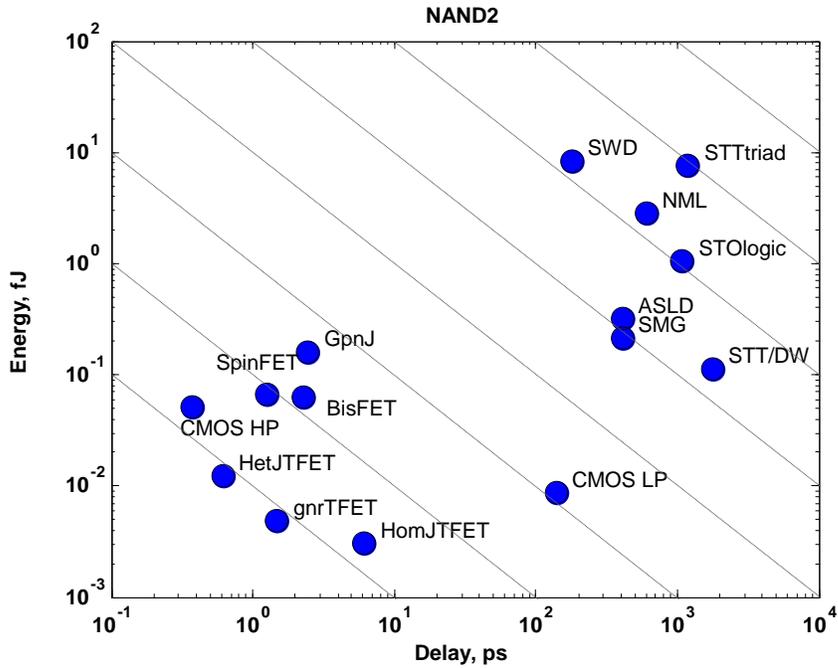

**Figure 54. Energy vs. delay of 2-input NAND gates. Spintronic devices use current-controlled switching with Vdd=0.01V. The preferred corner is bottom left.**

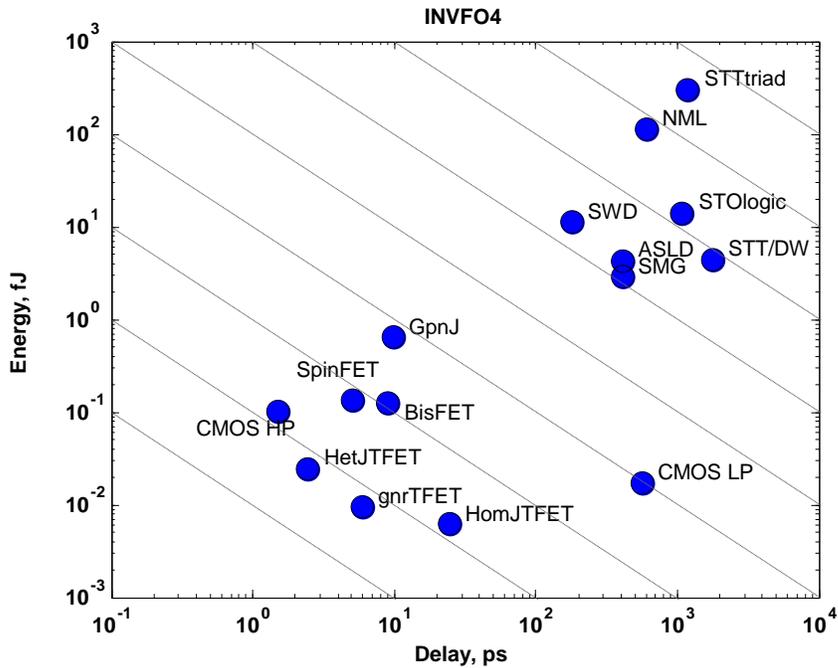

**Figure 55. Energy vs. delay of inverters with fanout of 4. Spintronic devices use current-controlled switching with Vdd=0.1V. The preferred corner is bottom left.**



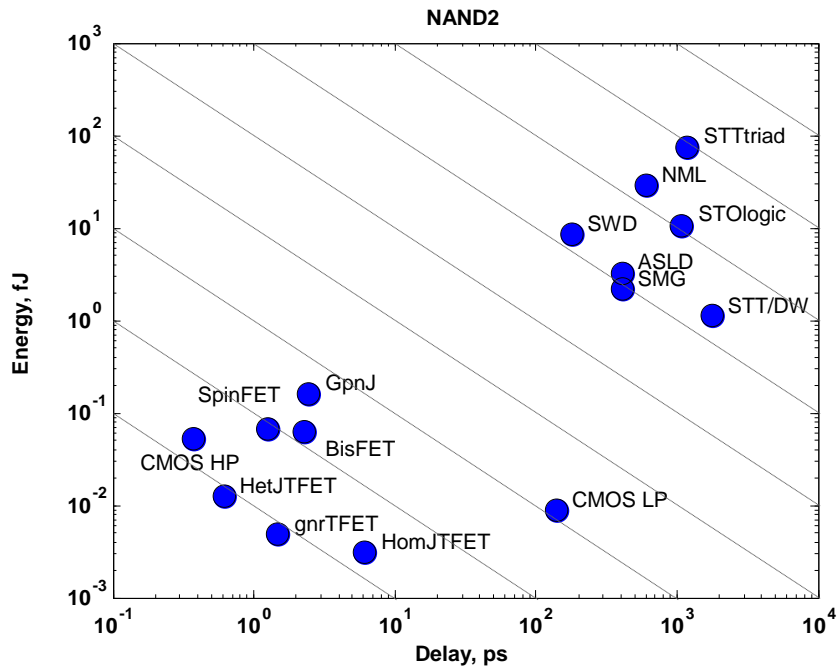

**Figure 56. Energy vs. delay of 2-input NAND gates. Spintronic devices use current-controlled switching with Vdd=0.1V. The preferred corner is bottom left.**

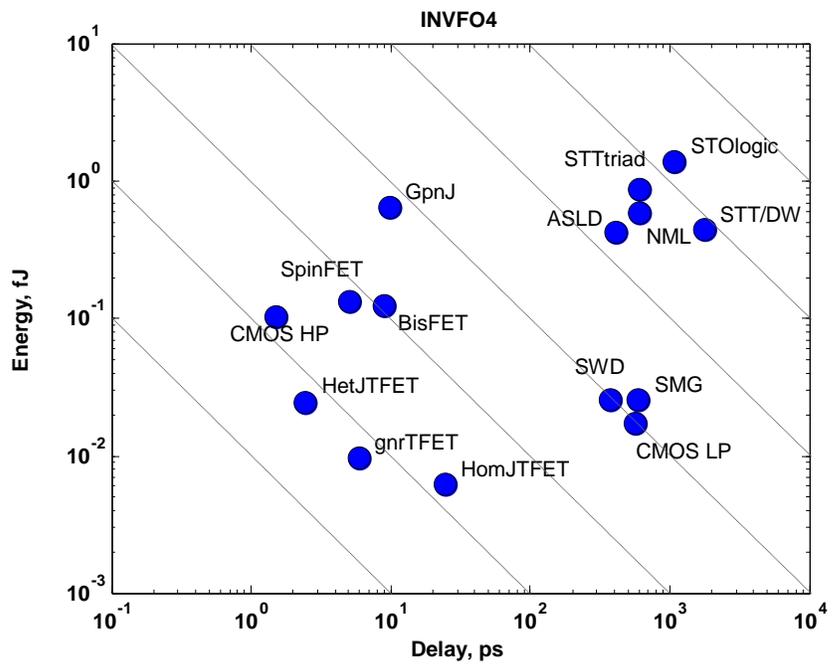

**Figure 57. Energy vs. delay of inverters with fanout of 4. Spintronic devices use multiferroic voltage controlled switching. The preferred corner is bottom left.**



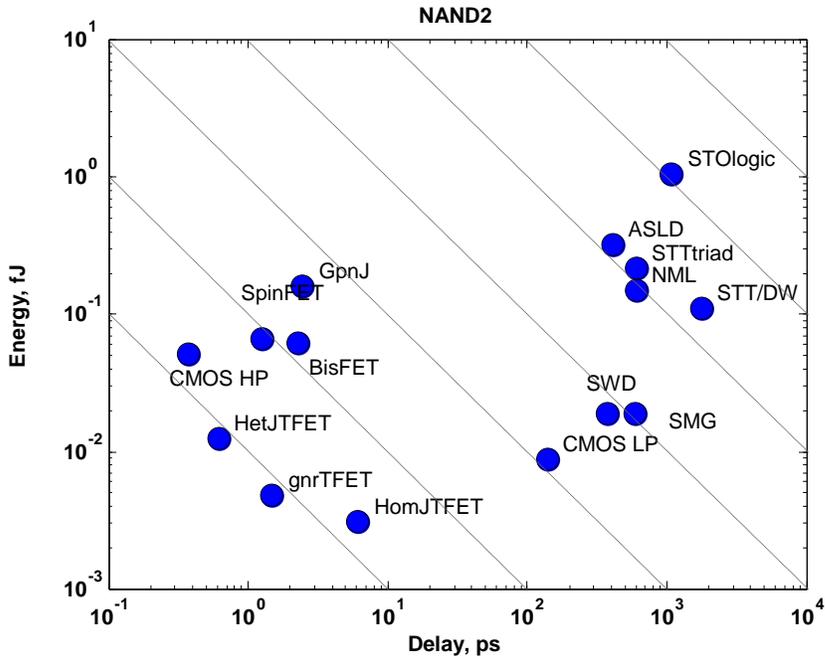

**Figure 58. Energy vs. delay of 2-input NAND gates. Spintronic devices use multiferroic voltage controlled switching. The preferred corner is bottom left.**

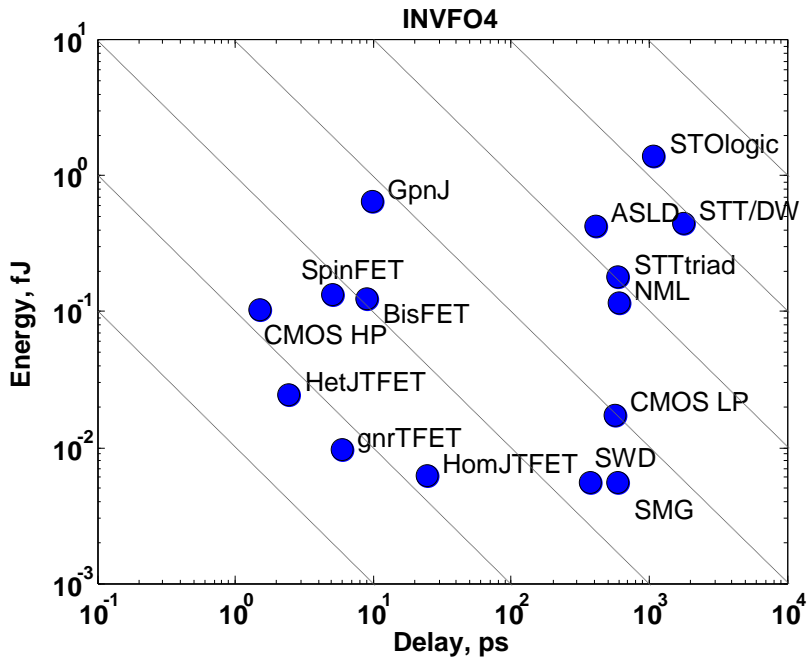

**Figure 59. Energy vs. delay of inverters with fanout of 4. Spintronic devices use magnetostrictive voltage controlled switching. The preferred corner is bottom left.**



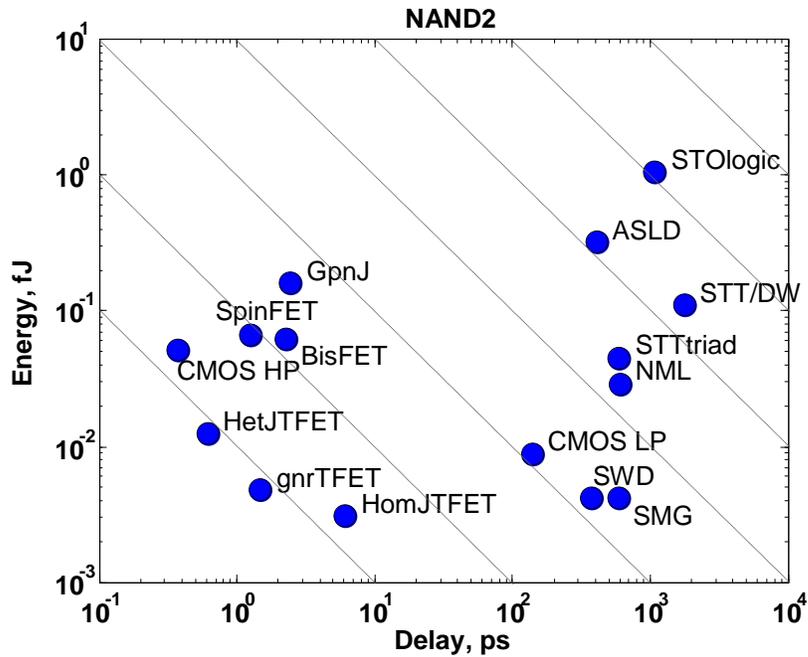

**Figure 60. Energy vs. delay of 2-input NAND gates. Spintronic devices use magnetostrictive voltage controlled switching. The preferred corner is bottom left.**

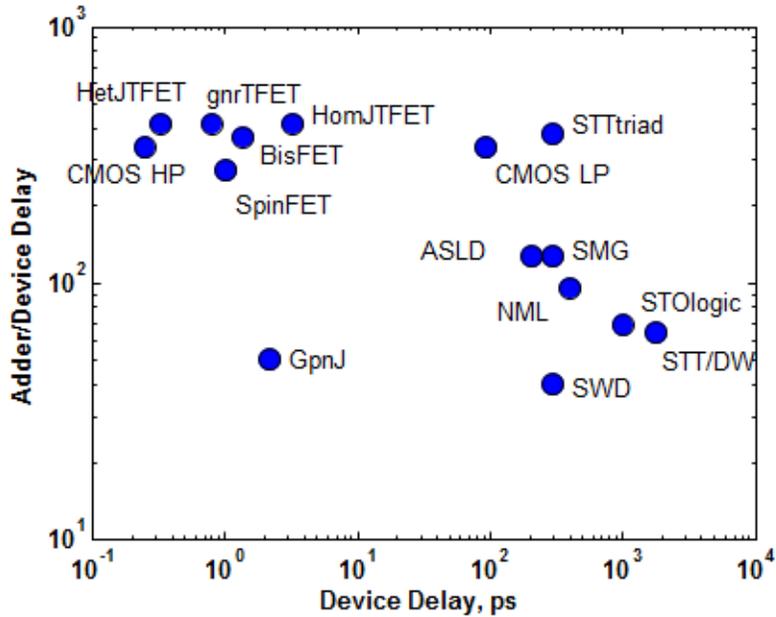

**Figure 61. Ratio of the delay of the 32-bit adders and their intrinsic device delay vs. intrinsic device.**



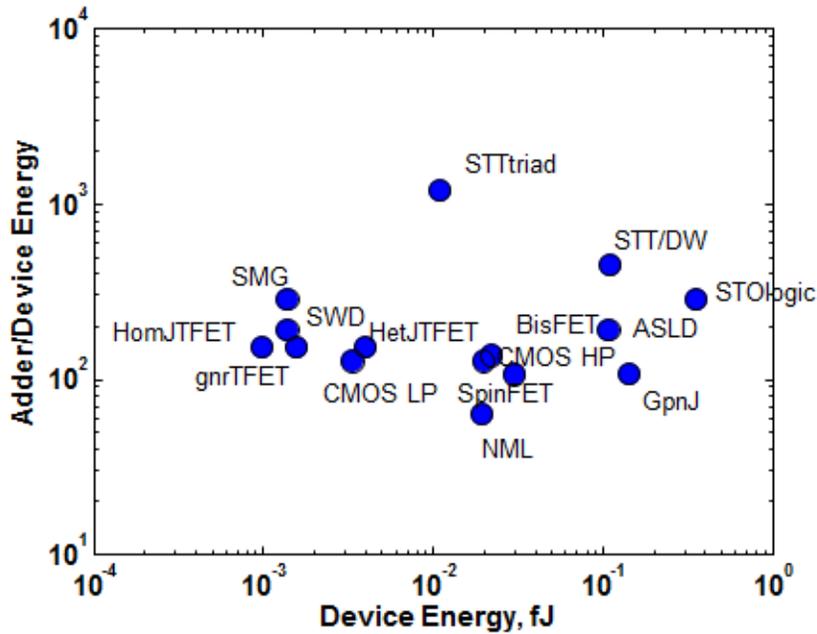

**Figure 62. . Ratio of switching energy of the 32-bit adders and their intrinsic device switching energy vs. intrinsic device switching energy.**

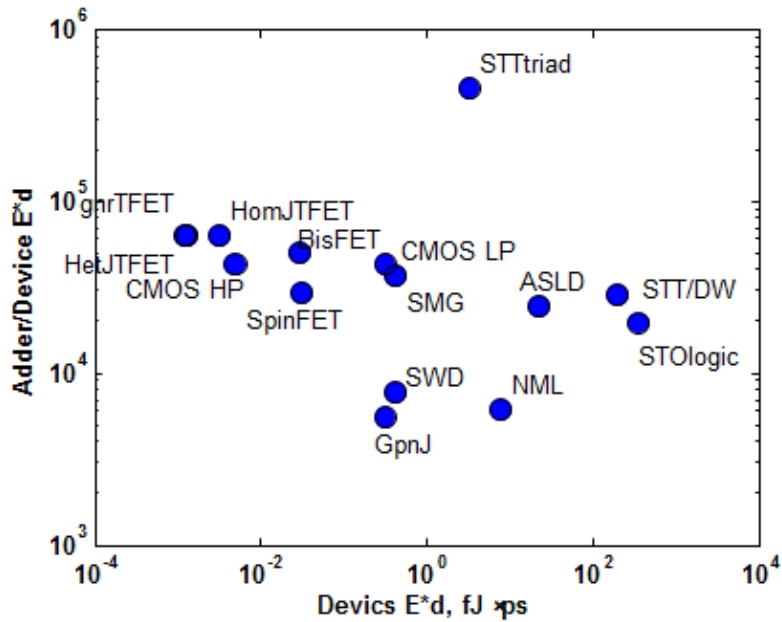

**Figure 63. Ratio of the energy-delay product of the 32-bit adders and their intrinsic device energy-delay product vs. intrinsic device energy-delay product .**



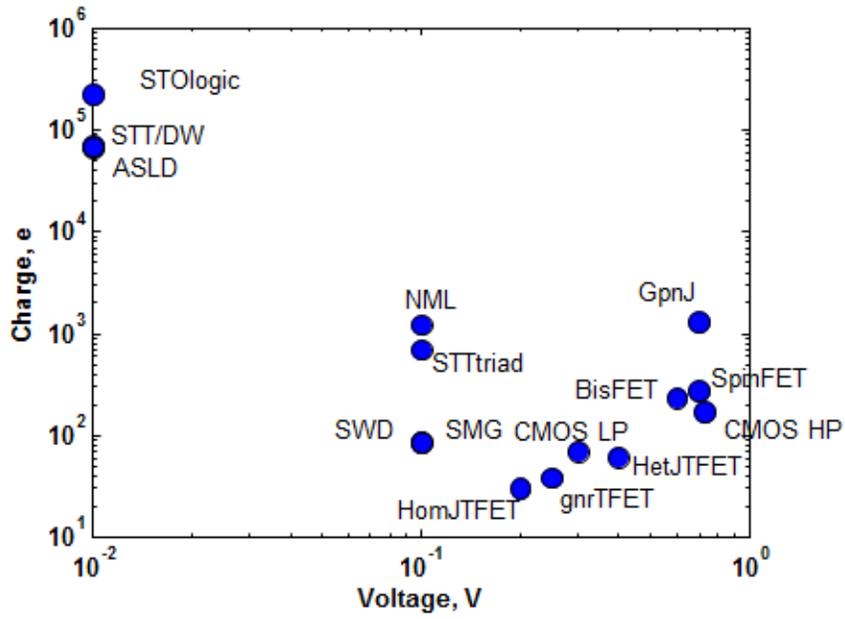

**Figure 64. Charge vs. voltage for intrinsic devices.**

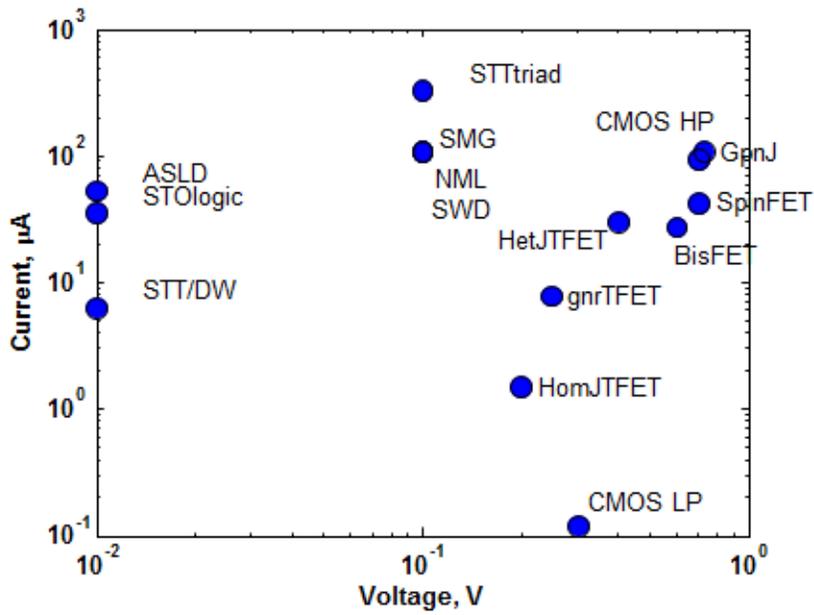

**Figure 65. Current vs. voltage for intrinsic devices.**



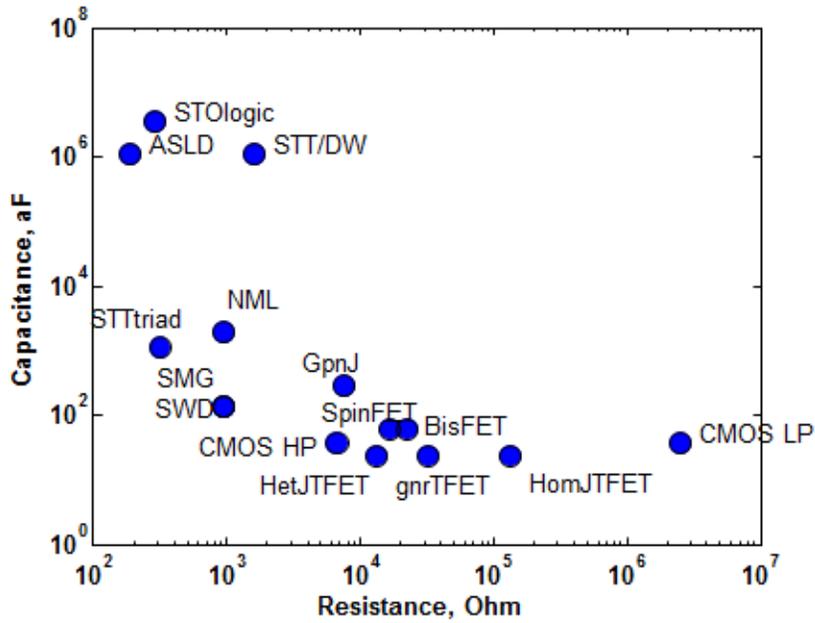

**Figure 66. Capacitance vs. resistance of intrinsic devices.**

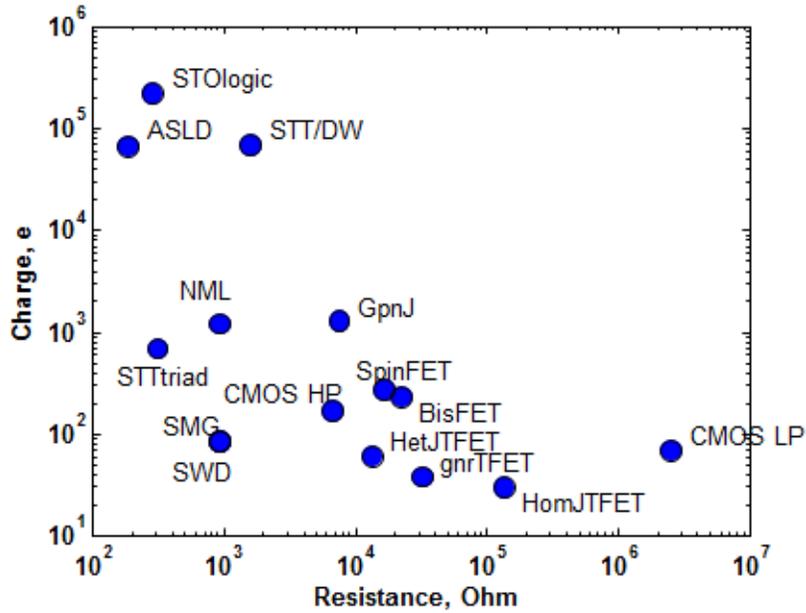

**Figure 67. Charge vs. resistance of intrinsic devices.**



## 18. References


[1] International Technology Roadmap for Semiconductors, available http://www.itrs.net/, Chapter PIDS (2011).

[2] V. V. Zhirnov, R. K. Cavin, J. A. Hutchby, and G. I. Bourianoff, "Limits to binary logic switch scaling - a gedanken model," Proceeding IEEE, vol. 91, no. 11, pp. 1934–1939, Nov. 2003.

[3] K. Bernstein, presentation, NRI Benchmarking Workshop, Aug. 2010. Also see K. Bernstein, R. K. Cavin III, W. Porod, A. Seabaugh, and J. Welser, "Device and Architecture Outlook for Beyond-CMOS Switches", Proceedings IEEE v. 98, pp. 2169-84 (2010).

[4] Available online: https://nanohub.org/tools/nribench/browser/trunk/src.

[5] K. Bernstein, presentation, NRI Annual Review. Nov. 2011.

[6] C. Reischer and D. A. Simovici, "On the implementation of set-valued non-Boolean switching functions", Proceedings of the 21st International Symposium on Multiple-Valued Logic, pp. 166-172 (1991).

[7] C. Mead, "Neuromorphic electronic systems", Proceedings of the IEEE, v. 78, pp. 1629 – 1636 (1990).

[8] "Nanoelectronics and Information Technology", ed. R. Waser (Wiley-VCH, Berlin, 2005).

[9] I. Zutic, J. Fabian, and S. Das Sarma, "Spintronics: Fundamentals and applications", Rev. Mod. Phys. 76, 323 (2004).





[10] D. E. Nikonov and G. I. Bourianoff, "Operation and Modeling of Semiconductor Spintronics Computing Devices", J. Supercond. and Novel Magnetism, v. 21, no. 8, pp. 479-493 (2008).

[11] U. E. Avci, R. Rios, K. Kuhn, I. A. Young, "Comparison of performance, switching energy and process variations for the TFET and MOSFET in logic", Proceedings of the 2011 VLSI Technology Symposium, pp. 124-125, June 2011.

[12] A. C. Seabaugh and Q. Zhang, "Low-Voltage Tunnel Transistors for Beyond-CMOS Logic", Proceedings IEEE, v. 98, pp. 2095-2010 (2010).

[13] H. Kroemer, "A proposed class of hetero-junction injection lasers", Proceedings of the IEEE, v. 51, pp. 1782 − 1783 (1963).

[14] M. Luisier and G. Klimeck, "Performance analysis of statistical samples of graphene nanoribbon tunneling transistors with line edge roughness", Appl. Phys. Lett. v. 94, 223505 (2009).

[15] S. O. Koswatta, M. S. Lundstrom, and D. E. Nikonov, "Performance Comparison Between p-i-n Tunneling Transistors and Conventional MOSFETs", IEEE Transactions on Electronic Devices v. 56, pp. 456-465 (2009).

[16] T. Low and J. Appenzeller, "Electronic transport properties of a tilted graphene p-n junction", Phys. Rev. B v.80, 155406 (2009).

[17] D. Reddy, L. F. Register, E. Tutuc, and Sanjay K. Banerjee, "Bilayer PseudoSpin Field-Effect Transistor (BiSFET): A Proposed New Logic Device", IEEE Trans. Electronic Devices, v. 30, pp. 158 − 160 (2009).

[18] H. Min, R. Bistritzer, J.-J. Su, and A. H. MacDonald, "Room-temperature





superfluidity in graphene bilayers," Phys. Rev. B, Condens. Matter, vol. 78, no. 12, p. 121401 (2008).

[19] M. Y. Kharitonov and K. B. Efetov, "Excitonic condensation in a double-layer graphene system" Semicond. Sci. Technol. v. 25, p. 034004 (2010).

[20] M. J. Gilbert, "Performance Characteristics of Scaled Bilayer Graphene Pseudospin Devices", IEEE T. Electron Devices, v. 75, pp. 3059-67 (2010).

[21] B. Dellabetta and M. J. Gilbert, "Performance characteristics of strongly correlated bilayer graphene for post-CMOS logic devices", Silicon Nanoelectronics Workshop (SNW), pp. 1-2 (2010).

[22] B. Dellabetta and M. J. Gilbert, Device Research Conference (DRC) digest, "Effect of disorder on superfluidity in double layer graphene", pp. 65 – 66 (2011).

[23] L. F. Register, X. Mou, D. Reddy, W. Jung, I. Sodeman, D. Pesin, A. Hassibi, A. H. MacDonald, and S. K. Banerjee, "Bilayer Pseudo-Spin Field Effect Transistor (BiSFET): Concepts and Critical Issues for Realization", ECS Transactions (2012).

[24] S. Sugahara and M. Tanaka, "A spin metal–oxide–semiconductor field-effect transistor using half-metallic-ferromagnet contacts for the source and drain," Applied Physics Letters, vol. 84, p. 2307 (2004).

[25] K. Galatsis, presentation, NRI Benchmarking Workshop, Aug. 2011.

[26] J. A. Currivan, Y. Jang, M. D. Mascaro, M. A. Baldo, and C. A. Ross, "Low Energy Magnetic Domain Wall Logic in Short, Narrow Ferromagnetic Wires", IEEE Mag. Lett. (2012), to be published.





[27] D. E. Nikonov and G. I. Bourianoff, "Recent progress, opportunities and challenges for beyond-CMOS information processing technologies", ECS Trans., v. 35, no. 2, pp. 43–53, May 2011.

[28] D. E. Nikonov, G.I. Bourianoff, and T. Ghani, "Proposal of a Spin Torque Majority Gate Logic", IEEE Electron. Device Lett., v. 32, n. 8, pp. 1128-30, Aug. 2011.

[29] A. Kozhanov and J. Allen, presentation, NRI Benchmarking Workshop, Aug. 2011.

[30] I. Krivorotov and D. Markovic, presentation, NRI Benchmarking Workshop, Aug. 2011.

[31] B. Behin-Aein, D. Datta, S. Salahuddin, and S. Datta, "Proposal for an all-spin logic device with built-in memory", Nature Nano v. 5, p. 266 (2010).

[32] B. Behin-Aein, A. Sarkar, S. Srinivasan and S. Datta, "Switching energy-delay of all spin logic devices", Appl. Phys. Lett. 98, 123510, 2011.

[33] C. Augustine, G. Panagopoulos, B. Behin-Aein, S. Srinivasan, A. Sarkar, and K. Roy, "Low-Power Functionality Enhanced Computation Architecture Using Spin-Based Devices", IEEE International Symposium on Nanoscale Architectures (NANOARCH), p. 129-136 (2011).

[34] A. Khitun and K. L. Wang, "Nano scale computational architectures with Spin Wave Bus", Superlatt. and Microstruct. 38, 184 (2005).

[35] A. Khitun, M. Bao, J.-Y. Lee, K. L. Wang, D. W. Lee, S. X. Wang, and I. V. Roshchin, "Inductively Coupled Circuits with Spin Wave Bus for Information Processing", J. of Nanoelectronics and Optoelectronics 3, 24 (2008).





36 R. P. Cowburn and M. E. Welland, "Room Temperature Magnetic Quantum Cellular Automata", Science 287, 1466 (2000).

37 S. Datta and B. Das, "Electronic analog of the electro optic modulator," Applied Physics Letters, vol. 56, pp. 665-667 (1990).

38 M. Kabir, D. Unluer, L. Li, A. W. Ghosh, M. R. Stan, "Electronic ratchet: A non-equilibrium, low power switch", 11[th] IEEE Conference on Nanotechnology, pp. 482-486 (2011).

39 Y. P. Chen and J. Hu, presentation, NRI Benchmarking Workshop, Aug. 2010.

40 V. Saripalli, V. Narayanan, and S. Datta, "Ultra Low Energy Binary Decision Diagram Circuits Using Few Electron Transistors", Nano-Net, Lecture Notes of the Institute for Computer Sciences, Social Informatics and Telecommunications Engineering, v. 20, Part 2, pp. 200-209, Springer, 2009.

41 H. Raza, E. C. Kan, T. Z. Raza, T.-H. Hou, "On the Possibility of an Electronic-structure Modulation Transistor", online http://arxiv.org/abs/0812.0123v2 (2009).

42 S. A. Wolf, J. Lu, M. R. Stan, E. Chen, and D. M. Treger, "The Promise of Nanomagnetics and Spintronics for Future Logic and Universal Memory", Proceedings IEEE, v. 98, pp. 2155-2168 (2010).

43 M. Kabir, M. R. Stan, S. A. Wolf, R. B. Comes, J. Lu, "RAMA: a self-assembled multiferroic magnetic QCA for low power systems", Proceedings of the 21st edition of the Great lakes symposium on VLSI, pp. 25-30, 2011.

44 H. Chen, L. F. Register, S. K. Banerjee, "Resonant Injection Enhanced Field Effect Transistor for Low Voltage Switching: Concept and Quantum Transport Simulation",





International Conference on Simulation of Semiconductor Processes and Devices (SISPAD '09), pp.1-4, 2009.

[45] D. A. Allwood, G. Xiong, C. C. Faulkner, D. Atkinson, D. Petit, and R. P. Cowburn, "Magnetic domain-wall logic", Science 309, 1688 (2005).

[46] D. E. Nikonov, G. I. Bourianoff, and P. A. Gargini, "Simulation of Highly Idealized, Atomic Scale Magnetic Quantum Cellular Automata Logic Circuits", Journal of Nanoelectronics and Optoelectronics 3, 3 (2008).

[47] D. M. Newns, J. A. Misewich, C. C. Tsuei, A Gupta, B. A. Scott, and A. Schrott, "Mott transition field effect transistor", Appl. Phys. Lett. 73, 780 (1998).

[48] J. Wunderlich,B.-G. Park, A. C. Irvine, L. P. Zârbo, E. Rozkotová, P. Nemec, V. Novák, J. Sinova, T. Jungwirth, "Spin Hall Effect Transistor", Science v. 330, pp. 1801-1804 (2010).

[49] P. M. Koenraad and M. E. Flatté, "Single dopants in semiconductors", Nature Materials, v. 10, pp. 91–100 (2011).

[50] A. Ney, C. Pampuch, R. Koch & K. H. Ploog, "Programmable computing with a single magnetoresistive element", Nature 425, 485 (2003).

[51] J. Wang, H. Meng, and J.-P. Wang, "Programmable spintronics logic device based on a magnetic tunnel junction element", J. Appl. Phys. 97, 10D509 (2005).

[52] S. Lee, S. Choa, S. Lee, and H. Shin, "Magneto-Logic Device Based on a Single-Layer Magnetic Tunnel Junction", IEEE Trans. Electron. Devices 54, 2040 (2007).





[53] S. Matsunaga, J. Hayakawa, S. Ikeda, K. Miura, H. Hasegawa, T. Endoh, H. Ohno, and T. Hanyu, "Fabrication of a Nonvolatile Full Adder Based on Logic-in-Memory Architecture Using Magnetic Tunnel Junctions", Appl. Phys. Express 1, 091301 (2008).

[54] A.A. High, A.T. Hammack, L.V. Butov, M. Hanson, A.C. Gossard, "Exciton optoelectronic transistor", Optics Letters 32, 2466 (2007).

[55] A.A. High, E.E. Novitskaya, L.V. Butov, M. Hanson, A.C. Gossard, "Control of exciton fluxes in excitonic integrated circuits", Science 321, 229 (2008).

[56] G. Grosso, J. Graves, A.T. Hammack, A.A. High, L.V. Butov, M. Hanson, A.C. Gossard, "Excitonic switches operating at around 100 K", Nature Photonics 3, 577 (2009).

[57] International Technology Roadmap for Semiconductors, available http://www.itrs.net/, Chapter ERD (2011).

[58] J. U. Lee, "Single Exciton Quantum Logic Circuits", IEEE J. Quant. Electronics, v. 48, pp. 1158-64 (2012).

[59] J. Wang, J. B. Neaton, H. Zheng, V. Nagarajan, S. B. Ogale, B. Liu, D. Viehland, V. Vaithyanathan, D. G. Schlom, U. V. Waghmare, N. A. Spaldin, K. M. Rabe, M. Wuttig, and R. Ramesh, "Epitaxial BiFeO3 Multiferroic Thin Film Heterostructures", Science, v. 299, pp. 1719-1722, March 2003.

[60] P. Shabadi, A. Khitun, K. Wong, P. K. Amiri, K. L. Wang, and C. A. Moritz, "Spin wave functions nanofabric update", Proc. IEEE/ACM Intl. Symposium on Nanoscale Architectures, pp. 107-113 (2011).





[61] T. Wu, A. Bur, P. Zhao, K. P. Mohanchandra, K. Wong, K. L. Wang, C. S. Lynch, and G. P. Carman, "Giant electric-field-induced reversible and permanent magnetization reorientation on magnetoelectric Ni/011[Pb(Mg1/3Nb2/3)O(1−x)–PbTiO3x heterostructure", Appl. Phys. Lett. 98, 012504 (2011).

[62] M. Fiebig, "Revival of the magnetoelectric effect", J. Phys. D: Appl. Phys. 38, R123–R152 (2005).

[63] J. A. Rabaey "Design Rules", available http://bwrc.eecs.berkeley.edu/classes/icbook/SLIDES/design_rules.ppt. Also see http://www.mosis.com/files/scmos/scmos.pdf.

[64] N. Weste and D. Harris, "CMOS VLSI Design: A Circuits and Systems Perspective", 4th ed. (Addison Wesley, 2010).

[65] C. Auth et al., "A 22nm High Performance and Low-Power CMOS Technology Featuring Fully-Depleted Tri-Gate Transistors, Self-Aligned Contacts and High Density MIM Capacitors", Symposium on VLSI Technology Digest of Technical Papers, pp. 131-132 (2012) .

[66] H. Liu, D. K. Mohata, A. Nidhi, V. Saripalli, V. Narayanan and S. Datta, "Exploration of Vertical MOSFET and Tunnel FET Device Architecture for Sub 10nm Node Applications",  Device Research Conference Digest, p. 233-234 (2012).

[67] C. Augustine, A. Raychowdhury, Y. Gao, M. Lundstrom, and K. Roy, "PETE: A device/circuit analysis framework for evaluation and comparison of charge based emerging devices", International Symposium on Quality of Electronic Design, p. 80-85 (2009).





[68] L. F. Register, D. Basu, and D. Reddy, "From Coherent States in Adjacent Graphene Layers toward Low-Power Logic Circuits", Advances in Condensed Matter Physics, v. 2011, 258731 (2011).

[69] C. Pan and A. Naeemi, private communication.

[70] G. L. Snider, A. O. Orlov, I. Amlani, X. Zuo, G. H. Bernstein, C. S. Lent, J. L. Merz, and W. Porod, "Quantum-dot cellular automata: Review and recent experiments (invited)", J. Appl. Phys. 85, 4283 (1999).

[71] Y. Gao, T. Low, and M. Lundstrom, "Possibilities for VDD = 0.1V Logic Using Carbon-Based Tunneling Field Effect Transistors", Symposium on VLSI Technology Digest of Technical Papers, pp. 180-181 (2009).

[72] J. A. Davis, V. K. De, and J. D. Meindl, "A Stochastic Wire-Length Distribution for Gigascale Integration (GSI)—Part II: Applications to Clock Frequency, Power Dissipation, and Chip Size Estimation", IEEE Trans. Electron Devices, v. 45, pp. 590-597 (1998).

[73] S. Tanachutiwat, J. U. Lee, and W. Wang, "Reconfigurable Multi-Function Logic Based on Graphene P-N Junctions", Proceedings of Design Automation Conference, pp. 883-888, Anaheim, California, USA, June 13-18, 2010.

[74] J. Z. Sun, "Spin angular momentum transfer in current-perpendicular nanomagnetic junctions", IBM J. Res. Dev. 50, 81-100 (2006).

[75] R. H. Koch, J. A. Katine, and J. Z. Sun, "Time-Resolved Reversal of Spin-Transfer Switching in a Nanomagnet", Phys. Rev. Lett., v. 92, 088302 (2004).





[76] S. Mangin, D. Ravelsona, J. A. Katine, M. J. Carey, B. D. Terris, and E. Fullerton, "Current-induced magnetization reversal in nanopillars with perpendicular anisotropy", nature Materials, 5, 210-215 (2006).

[77] M. F. Toney, E. E. Marinero, M. F. Doerner, and P. M. Rice, "High anisotropy CoPtCrB magnetic recording media", J. Appl. Phys. v. 94, pp. 4018-4023 (2003).

[78] J. T. Heron, M. Trassin, K. Ashraf, M. Gajek, Q. He, S.Y. Yang, D. E. Nikonov, Y-H. Chu, S. Salahuddin, and R. Ramesh, "Electric-Field-Induced Magnetization Reversal in a Ferromagnet-Multiferroic Heterostructure", Phys. Rev. Lett. 107, 217202 (2011).

[79] T. Maruyama, Y. Shiota, T. Nozaki, K. Ohta, N. Toda, M. Mizuguchi, A. A. Tulapurkar, and T. Shinjo, M. Shiraishi, S. Mizukami, Y. Ando and Y. Suzuki, "Large voltage-induced magnetic anisotropy change in a few atomic layers of iron", Nature Nano. 4, pp. 158-161 (2009).

[80] A. V. Khvalkovskiy, K. A. Zvezdin, Y. V. Gorbunov, V. Cros, J. Grollier, A. Fert, and A. K. Zvezdin, "High domain wall velocities due to spin currents perpendicular to the plane," Phys. Rev. Lett., vol. 102, no. 6, p. 067 206, Feb. 2009.

[81] C. A. Ross, INDEX review presentation, 2010.

[82] S. V. Pietambaram, N. D. Rizzo, R. W. Dave, J. Goggin, K. Smith, J. M. Slaughter, and S. Tehrani, "Low-power switching in magnetoresistive random access memory bits using enhanced permeability dielectric films ", Appl. Phys. Lett. 90, 143510 (2007).

[83] A. Imre, G. Csaba, L. Ji, A. Orlov, G. H. Bernstein, W. Porod, "Majority logic gate for magnetic quantum-dot cellular automata," Science, v. 311, pp. 205-208, (2006).





[84] Th. Gerrits, H. A. M. van den Berg, J. Hohlfeld, L. Bar, Th. Rasing, Nature 418, 509 (2002).

[85] D. B. Carlton, N. C. Emley, E. Tuchfeld, and J. Bokor, "Simulation Studies of Nanomagnet-Based Logic Architecture", Nano Lett., v. 8 , pp. 4173–4178 (2008).

[86] F. M. Spedalieri, A. P. Jacob, D. E. Nikonov, V. P. Roychowdhury, "Performance of Magnetic Quantum Cellular Automata and Limitations Due to Thermal Noise", IEEE Trans. Nano., v.10, n. 5, pp. 537-546 (2011).

[87] M. Niemier, private communication.

[88] K. Bernstein, presentation, NRI Benchmarking Workshop, Aug. 2011.